\documentclass[10pt,twocolumn,showpacs,preprintnumbers,amsmath,amssymb,aps,prb,longbibliography,superscriptaddress,footinbib]{revtex4-2}

\usepackage{xcolor}
\usepackage{diagbox}
\usepackage{graphicx}
\usepackage{dcolumn}
\usepackage{bm}

\usepackage[unicode=true,
bookmarks=true,bookmarksnumbered=true,bookmarksopen=false,
breaklinks=true,colorlinks=true]{hyperref}
\hypersetup{citecolor={blue},urlcolor={magenta}}

\begin{document}

\title{Bootstrapping Classical Systems with Quenched Disorder}

\author{Minjae Cho}
 \email{cho7@uchicago.edu}
\affiliation{Leinweber Institute for Theoretical Physics, University of Chicago, Chicago, IL 60637, USA}

\author{Yaprak Önder}
\affiliation{Department of Physics, Harvard University, Cambridge, Massachusetts 02138, USA}

\author{Eslam Khalaf}
\affiliation{Department of Physics, Harvard University, Cambridge, Massachusetts 02138, USA}

\author{Michael G. Scheer}
\affiliation{Department of Physics, Harvard University, Cambridge, Massachusetts 02138, USA}

\begin{abstract}

We introduce a bootstrap method for deriving bounds on disorder-averaged observables in classical statistical systems with quenched disorder. The three main ingredients of the bootstrap are the positivity of ensembles of probability measures, disorder-averaged equations of motion, and explicit expectation values of random coupling variables. Together, these ingredients lead to a linear programming problem over the space of disorder averages. By further employing the positivity of the two-replica Gram matrix, we also formulate a semidefinite programming problem that produces bounds on two-replica correlators. We demonstrate the method using two examples: the two-dimensional random site-diluted Ising model and the one-dimensional disordered contact process. For the latter example, the bootstrap bounds exhibit a region of multiple kinks that we relate to the Griffiths region, whose finite-size scaling analysis yields a typical correlation length exponent in good numerical agreement with the known value.
\end{abstract}

\maketitle

\section{\label{sec:introduction}Introduction}
Phases of matter constitute one of the most fundamental subjects in physics. The study of equilibrium phases has led to the development of many of the fundamental building blocks of modern physics, while nonequilibrium phases exhibit novel phenomena that are not observed in equilibrium settings. In both equilibrium and nonequilibrium systems, it is natural to ask what happens when a system is subject to random impurities that generate quenched disorder. Such extensions are not only realistic but also give rise to physics that is genuinely distinct from that of the clean case, i.e., the case without disorder.

There are several well-established results concerning systems with quenched disorder, including Anderson localization \cite{1958PhRv..109.1492A}, the Harris criterion \cite{ABHarris_1974}, the Imry-Ma argument \cite{PhysRevLett.35.1399}, and the Aizenman-Wehr theorem \cite{PhysRevLett.62.2503,Aizenman1990}. At the same time, several important questions remain unsettled, such as the nature of the spin-glass phase transition \cite{SFEdwards_1975,PhysRevLett.35.1792,Mezard1987}. Numerical Monte Carlo simulations of such systems face difficulties for structural reasons, including the need to average over numerous disorder realizations \cite{Jain1992}, complex energy landscapes, and Griffiths regions \cite{PhysRevLett.23.17}. Furthermore, the results are always obtained for finite systems and extrapolation to the thermodynamic limit is required. It is therefore desirable to develop alternative methods for studying disordered systems that can overcome these difficulties.

Motivated by these considerations, we develop a bootstrap framework for classical statistical systems with quenched disorder. Such a framework has already been established for clean equilibrium and nonequilibrium statistical systems \cite{Cho:2022lcj,Cho:2023ulr,Cho:2025dgc} and admits a natural extension to systems with quenched disorder. As we explain in detail throughout the remainder of this work, the framework reformulates the defining properties of the disorder average as a linear programming (LP) problem that produces bounds on disorder-averaged observables. These bounds are rigorous and apply to systems on strictly infinite lattices. We further extend the bootstrap framework to derive bounds on two-replica correlators by incorporating the positivity of the two-replica Gram matrix, leading to a semidefinite programming (SDP) problem. In \cite{unpubl1}, we also report a similar bootstrap formulation for quantum systems with quenched disorder. We remark that optimization methods have already been applied to disordered systems in various forms \cite{LBieche_1980,FBarahona_1982,Barahona1989,Goemans1995ImprovedAA}, but, to the best of our knowledge, the specific formulation presented in this work has not previously appeared.

We begin by briefly summarizing the main ideas underlying the disordered bootstrap framework. Consider a statistical system with configurations collectively denoted by $s$, such as spin configurations. We focus on systems described by probability measures defined over such configurations \footnote{This excludes systems described by signed or complex measures that appear in certain lattice gauge theories, for which positivity (\ref{eq:positivity}) does not apply.}, such as Gibbs measures. We also introduce the collective label $r$ for the couplings of the system, such as bond strengths, site dilution, external magnetic fields, and infection rates. For a fixed set of coupling values $r$, the expectation value of a function $g(s)$ is denoted by $\langle g(s)\rangle_r$. When the system is subject to quenched disorder, the couplings $r$ are drawn from a random distribution described by a probability measure $d\mu(r)$. We denote the corresponding average of a function $h(r)$ by an overline, $\overline{h(r)}\equiv\int d\mu(r)h(r)$. The disorder average of a function $f(s,r)$ is then defined as
\begin{equation}\label{eq:disorderAvg}
\overline{\langle f(s,r)\rangle}\equiv\overline{\langle f(s,r)\rangle_r}=\int d\mu(r) \langle f(s,r)\rangle_r.
\end{equation}

Our main observation is that the bootstrap framework allows us to treat the configuration variables $s$ and the random couplings $r$ on equal footing. In clean system bootstrap, we optimize over expectation values subject to consistency conditions that are necessarily satisfied whenever these expectation values correspond to an equilibrium Gibbs state or a nonequilibrium stationary state. Here, we will generalize such consistency conditions to include joint expectation values over both the state and the disorder probability distribution, as we describe in detail below \footnote{We also remark that a closely related idea of considering the joint space of random couplings and quantum states has been applied to the tensor network representation of quantum many-body states for quantum systems with quenched disorder in \cite{PhysRevLett.95.140501,SciPostPhys.6.3.031,Vervoort:2025auj}.}.

First, the disorder average of any nonnegative function $Q(s,r)$ of $s$ and $r$ must be nonnegative:
\begin{equation}\label{eq:positivity}
\text{Positivity: }\overline{\langle Q(s,r)\rangle}\geq0,~~\text{if }Q(s,r)\geq0~\text{for all }s,r.
\end{equation}
For a fixed disorder realization $r$, we are interested in either the equilibrium measures or the nonequilibrium invariant measures of the system. In many cases of interest, expectation values $\langle\cdots\rangle_r$ of such measures satisfy linear equations. Examples include the Dobrushin-Lanford-Ruelle (DLR) equations (also called the detailed balance equations) in equilibrium \cite{doi:10.1137/1113026,Dobrushin1968,Lanford1969} and stationary-state equations (also called the global balance equations) for nonequilibrium. These equations take the form $\langle q_\alpha(s,r)\rangle_r=0$ for an appropriate set of functions $q_\alpha(s,r)$ labeled by an index $\alpha$, as in the DLR equations (\ref{eq:DLRfix}) and invariance equations (\ref{eq:CPrinv}) discussed later. They hold for every disorder realization $r$, and we thus obtain
\begin{equation}\label{eq:linearEq}
\text{Equations: }\overline{\langle w(r)q_\alpha(s,r)\rangle}=0,~~\text{for any }w(r).
\end{equation}
Finally, the disorder average of a function $y(r)$ that is independent of $s$ is completely determined by the measure $d\mu(r)$:
\begin{equation}\label{eq:disorderValue}
\text{Disorder distribution: }\overline{\langle y(r)\rangle}=\int d\mu(r)y(r),
\end{equation}
where the right-hand side can be computed explicitly.

In fact, the combination of the three statements above is nothing more than the definition of disorder averages. In the bootstrap framework, these statements are treated as constraints on the space of disorder averages $\overline{\langle f(s,r)\rangle}$. Thus, The key step in bootstrapping disordered systems is to consider functions that depend on both $s$ and $r$, whose expectation values are subject to (\ref{eq:disorderValue}) which encodes the details of the disorder distribution, (\ref{eq:linearEq}) which promotes the DLR or stationary-state equations to their disordered forms, and (\ref{eq:positivity}) which encodes the positivity of the measure over the product space.

By minimizing or maximizing a specific disorder average of interest subject to these constraints, we can obtain a bootstrap lower or upper bound on it. The constraints (\ref{eq:linearEq}) and (\ref{eq:disorderValue}) are linear equalities among disorder averages. In contrast, the positivity constraints (\ref{eq:positivity}) are inequalities among them and may be expressed either as linear inequalities or as positive semidefinite (PSD) matrix constraints, depending on the details of the system. These constraints then give rise to standard LP or SDP problems, respectively, making the computation of bootstrap bounds straightforward using standard solvers such as MOSEK \cite{mosek2026}.

In this work, we apply the bootstrap method to two examples. The first is the equilibrium random site-diluted Ising model (RSIM) \cite{PhysRev.115.824,PhysRevLett.5.366} in two dimensions. This is a disordered variant of the ordinary Ising model in which the spin at each site can be independently and randomly diluted by being disconnected from the rest of the system. The ferromagnetic phase transition of the clean system may or may not exist, depending on the probability of dilution. Using the disordered bootstrap, we obtain bounds on the disorder-averaged magnetization and nearest-neighbor correlator for this system. For this example, we also show how to obtain bootstrap bounds on two-replica correlators.

The second example is the nonequilibrium disordered contact process (DCP) \cite{10.1214/aop/1176990331} in one dimension, in which the infection rate at each site is independently and randomly chosen from a set of values. The clean contact process \cite{10.1214/aop/1176996493,liggett1985interacting} famously exhibits an absorbing phase transition that belongs to the directed percolation universality class \cite{Janssen1981}, while its disordered counterpart is believed to belong to the same universality class as the one-dimensional quantum Ising spin chain with a random transverse field \cite{PhysRevLett.90.100601}, an example that we bootstrap in \cite{unpubl1}. We obtain upper bounds on the disorder-averaged infection density similarly to the clean case discussed in \cite{Cho:2025dgc}.

Both models possess a so-called Griffiths region \cite{PhysRevLett.23.17} between the critical point of the clean system and the infinite-randomness fixed point. In this range of couplings, rare spatial regions may occur in which ferromagnetism or infection persists even though the global system is paramagnetic or inactive. The bootstrap bounds on the order parameters of these systems, presented in the main text, exhibit curious multiple-kink features that are absent in the clean case. We later argue that these features are related to the Griffiths region. This argument is supported by a finite-size scaling analysis that extracts the typical correlation length exponent of the one-dimensional DCP, whose value is in good agreement with the known result.

\section{MODELS}
We first review the two models, the RSIM and the DCP, to which the bootstrap method will be applied. A careful examination of their definitions and properties naturally leads to the bootstrap framework. In particular, we formulate these definitions and properties entirely in terms of disorder averages, which will serve as the bootstrap variables to be bounded. We consider both models on the infinite $\mathbb{Z}^d$ lattice. At each site $i\in\mathbb{Z}^d$, there is a spin degree of freedom $s_i\in\{-1,1\}$ as well as a random coupling $r_i$ drawn from a specified disorder measure $d\mu(r)$. The random couplings are independent and identically distributed (i.i.d.) across lattice sites. In this work, we focus on the case in which $r_i$ takes values in $\{-1,1\}$. Generalization to an arbitrary finite set is straightforward in principle.

In both models, we are interested in probability measures over the product space of spins and couplings, whose expectation values are disorder averages. The statement that a distribution over this product space is a positive measure can be expressed in terms of disorder averages using indicator functions. Given subsets $A,B\subset\mathbb{Z}^d$, fixed spin assignments $u_i\in\{-1,1\}$ for $i\in A$, and fixed coupling assignments $v_j\in\{-1,1\}$ for $j\in B$, the corresponding indicator function is defined as
\begin{eqnarray}\label{eq:indicator}
        {}&&I_{A,B;u,v}(s,r)\equiv\prod_{i\in A}{1+u_is_i\over2}\prod_{j\in B}{1+v_jr_j\over2}\nonumber
        \\
        &&=
    \begin{cases}
        1 & \text{if } s_i=u_i, ~\forall i\in A~\text{and}~r_j=v_j,~\forall j\in B\\
        0   & \text{otherwise.}
    \end{cases}
\end{eqnarray}
Since this function is always nonnegative, (\ref{eq:positivity}) implies
\begin{equation}\label{eq:indPos}
    \overline{\langle  I_{A,B;u,v}(s,r) \rangle}\geq0.
\end{equation}
In fact, this condition is not only necessary but also sufficient, in the sense that any distribution whose expectation values satisfy (\ref{eq:indPos}) for all $A,B,u,v$ and obey linearity must be a positive measure \cite{Cho:2023ulr}. It becomes a probability measure if the normalization condition
\begin{equation}\label{eq:norm}
    \overline{\langle1\rangle}=1
\end{equation}
is also satisfied. $\overline{\langle  I_{A,B;u,v}(s,r) \rangle}$ is then the probability of the event $s_i=u_i, ~\forall i\in A~\text{and}~r_j=v_j,~\forall j\in B$ occurring. Of course, the RSIM and the DCP are not defined by arbitrary probability measures on the product space of spins and couplings. The equilibrium and invariant measures of interest must satisfy additional properties, which we now discuss.

\subsection{Equilibrium example: RSIM}
The Hamiltonian for the RSIM is given by
\begin{equation}\label{eq:RSIMhamil}
    H(s,r)=-\sum_{(i,j)}n_in_js_is_j,
\end{equation}
where the sum runs over all nearest-neighbor pairs $(i,j)$, and $n_i={1+r_i\over2}\in\{0,1\}$ is the random coupling representing site dilution. When $n_i=0$, the spin $s_i$ completely decouples from the rest of the system. The quenched disorder is parameterized by a number $p\in[0,1]$, such that $n_i$ is $1$ with probability $p$ and $0$ with probability $1-p$, with the values chosen independently across the lattice sites. The case $p=1$ corresponds to the clean system, namely, the ordinary Ising model, which exhibits a ferromagnetic phase transition for $d\geq2$. For $d=2$, the critical value of the inverse temperature $\beta$ in the clean case is $\beta_0={\log(1+\sqrt{2})\over2}$ \cite{PhysRev.60.252,PhysRev.60.263}, so that spontaneous magnetization develops for $\beta\geq\beta_0$. In contrast, for $p\leq p_c\approx0.592746$ \cite{2026arXiv260321303B}, site dilution is sufficiently strong that ferromagnetism is absent at any temperature. In the intermediate range $p_c<p<1$, a ferromagnetic phase transition occurs at a critical value $\beta=\beta_c(p)$, corresponding to an infinite-randomness fixed point, with $\beta_0\leq\beta_c(p)$. The region $\beta_0\leq\beta\leq\beta_c(p)$ is called the Griffiths region, in which disorder may produce rare regions with nontrivial local magnetization that takes an exponentially long time to relax to the globally paramagnetic state.

We make one remark concerning the observables of interest. In the presence of site dilution, natural observables are disorder averages of local spins multiplied by the dilution variables. For example, assuming translation invariance, the disordered magnetization is given by $\overline{\langle n_is_i \rangle}$. For any fixed disorder realization $r$, however, we have $\langle \prod_{i\in A}n_i s_i\rangle_r=\langle \prod_{i\in A} s_i\rangle_r$. When $n_i=1$, the two sides are trivially equal, while when $n_i=0$, the spin $s_i$ completely decouples from the rest of the system, so averaging over $s_i=\pm1$ gives zero, and both sides vanish. Therefore, $\overline{\langle\prod_{i\in A}n_is_i\rangle}=\overline{\langle\prod_{i\in A}s_i\rangle}$, and the disordered magnetization, for example, can equivalently be written as $\overline{\langle s_i \rangle}$.

\subsubsection{DLR equations}
For a fixed disorder realization $r$, the corresponding equilibrium measure is given by the Gibbs measure ${e^{-\beta H(s,r)}\over \sum_s e^{-\beta H(s,r)}}$. On a strictly infinite lattice, however, this expression is ill-defined. A proper definition of the Gibbs measure is provided by the DLR equations \cite{doi:10.1137/1113026,Dobrushin1968,Lanford1969} for the disorder-fixed expectation values $\langle\cdots\rangle_r$. These equations require the following conditional probability to satisfy
\begin{eqnarray}\label{eq:DLRfix}
    {}&&P(s_i=u_i|s_k=u_k,k\in W)={\langle {1+u_is_i\over2}\prod_{k\in W}{1+u_ks_k\over2} \rangle_r\over\langle\prod_{k\in W}{1+u_ks_k\over2}\rangle_r}\nonumber
    \\
    &&~~~~~~~~~~~=\left[ 1+\exp\left(-2\beta r_iu_i\sum_{l\in N(i)}r_lu_l\right) \right]^{-1},
\end{eqnarray}
for any site $i\in\mathbb{Z}^d$, any subset $W\subset\mathbb{Z}^d$ that contains the set $N(i)$ of nearest neighbors of $i$ but does not contain $i$ itself, and any spin assignments $u_i$ and $u_k$ for $k\in W$. The left-hand side of (\ref{eq:DLRfix}) is the probability that $s_i=u_i$ conditioned on $s_k=u_k$ for $k\in W$. It can therefore be expressed as the ratio of two expectation values of indicator functions, as shown in the first line. Note that the second line is a fixed number once $\beta,r_i,r_l,u_i$ and $u_l$ are specified. The DLR equations are therefore linear equations among the expectation values $\langle\cdots\rangle_r$, and any probability measure satisfying (\ref{eq:DLRfix}) is a Gibbs measure, which may or may not be unique. These are the equations $\langle q_\alpha(s,r)\rangle_r=0$ discussed above in connection with (\ref{eq:linearEq}), where $\alpha$ collectively labels the choices of $i,W,u_i$ and $u_l$.

Following (\ref{eq:linearEq}), we promote the DLR equations (\ref{eq:DLRfix}) to the disorder-averaged DLR equations
\begin{eqnarray}
    {}&&\overline{\bigg\langle w(r){1+u_is_i\over2}\prod_{k\in W}{1+u_ks_k\over2} \bigg\rangle}
    \\
    &&=\overline{\bigg\langle w(r)\prod_{k\in W}{1+u_ks_k\over2}\left[ {\scriptstyle1+\exp\left(-2\beta r_iu_i\sum_{l\in N(i)}r_lu_l\right)} \right]^{-1}\bigg\rangle},\nonumber
\end{eqnarray}
which holds for any function $w(r)$. By choosing the function $w(r)$ to be the indicator function for $r_j=v_j$ over $j\in B\subset\mathbb{Z}^d$, we arrive at
\begin{eqnarray}\label{eq:DLRdis}
    {}&&\overline{\bigg\langle I_{W\cup\{i\},B;u,v}(s,r)\bigg\rangle}
    \\
    &&=\overline{\bigg\langle I_{W,B;u,v}(s,r)\left[ 1+\exp\left(-2\beta r_iu_i\sum_{l\in N(i)}r_lu_l\right) \right]^{-1}\bigg\rangle}.\nonumber
\end{eqnarray}
Thus, (\ref{eq:DLRdis}) must hold for any choices of $v_j$ and $B$, in addition to all the choices already specified for (\ref{eq:DLRfix}). Any probability measure on the product space of spin and site-dilution configurations whose expectation values satisfy (\ref{eq:DLRdis}) defines the RSIM.

Note that the expression $[1+\exp(\cdots)]^{-1}$ in (\ref{eq:DLRdis}) can be rewritten as a polynomial in $r_i$ and $r_l$ because they are binary variables. Therefore, the disorder-averaged DLR equations are linear equations among disorder averages of the indicator functions $I_{A,B;u,v}(s,r)$ defined in (\ref{eq:indicator}). In particular, if $B$ contains $N(i)\cup\{i\}$, then $r_i$ and $r_l$ can be replaced by $v_i$ and $v_l$, respectively, due to the indicator, resulting in
\begin{eqnarray}\label{eq:DLRdisBW}
    {}&&\overline{\bigg\langle I_{W\cup\{i\},B;u,v}(s,r)\bigg\rangle}
    \\
    &&=\overline{\bigg\langle I_{W,B;u,v}(s,r) \bigg\rangle}\left[ 1+\exp\left(-2\beta v_iu_i\sum_{l\in N(i)}v_lu_l\right) \right]^{-1},\nonumber
\end{eqnarray}
where the $[1+\exp(\cdots)]^{-1}$ term has been factored out of the disorder average.

\subsubsection{Disorder distribution}
The variables $r_i$ are i.i.d. over $\mathbb{Z}^d$. Each variable is $1$ with probability $p$ and $-1$ with probability $1-p$. This information completely determines the disorder average of any function that depends only on $r$ and not on $s$. In particular,
\begin{equation}\label{eq:disorderDistR}
    \overline{\bigg\langle\prod_{j\in B}r_j\bigg\rangle}=(2p-1)^{|B|},~~\forall B\subset\mathbb{Z}^d,
\end{equation}
where $|B|$ is the number of sites in the subset $B$. These relations provide additional linear equations for the disorder averages and encode information about the disorder distribution.

\subsubsection{Griffiths inequalities}
The Hamiltonian (\ref{eq:RSIMhamil}) is ferromagnetic in the sense that the coupling $n_in_j$ is nonnegative for every fixed disorder realization. For such systems at a fixed disorder realization, there is a series of inequalities known as the Griffiths inequalities \cite{10.1063/1.1705219,10.1063/1.1705220,Griffiths1967}, and there exists at least one extremal Gibbs measure that satisfies them. We focus on the first Griffiths inequalities, which are linear in $\langle\cdots\rangle_r$. The other Griffiths inequalities are nonlinear and could, in principle, admit convex relaxations amenable to the bootstrap, but we do not pursue this direction in this work. For any fixed disorder realization $r$, the first Griffiths inequalities read
\begin{equation}
    \bigg\langle \prod_{i\in A} s_i\bigg\rangle_r\geq0,~~\forall A\subset\mathbb{Z}^d.
\end{equation}
Therefore, if we multiply this expression by any positive function of $r$ and then take the disorder average, the result remains nonnegative. Choosing the positive functions to be indicator functions of $r$, we obtain the disorder-averaged first Griffiths inequalities
\begin{equation}\label{eq:G1}
    \overline{\bigg\langle \left(\prod_{j\in B}{1+v_jr_j\over2}\right)\prod_{i\in A}s_i \bigg\rangle}\geq0,
\end{equation}
which must hold for any subset $B\subset\mathbb{Z}^d$ and any disorder assignments $v_j$ on it.

\subsubsection{Symmetries}
The Hamiltonian (\ref{eq:RSIMhamil}) is invariant under the spatial symmetries of $\mathbb{Z}^d$ and under the usual global spin-flip $\mathbb{Z}_2$ symmetry, which acts on spin configurations as $s_i\rightarrow-s_i$ for all $i\in \mathbb{Z}^d$ simultaneously. Therefore, there exists a probability measure on the product space that satisfies the disorder-averaged DLR equations (\ref{eq:DLRdis}) and is invariant under these symmetries.

For the case $d=2$, the spatial symmetry group $S$ is the semidirect product of the translation group $\mathbb{Z}^2$ and the dihedral group $D_4$: $S=\mathbb{Z}^2 \rtimes D_4$. Given a group element $g\in S$, we denote its action on a lattice subset $A\subset\mathbb{Z}^2$ by $g(A)$. The disorder averages of probability measures that respect $S$ then satisfy
\begin{equation}\label{eq:spatialsym}
    \overline{\bigg\langle\prod_{i\in A} s_i\prod_{j\in B}r_j\bigg\rangle}=\overline{\bigg\langle\prod_{i\in g(A)} s_i\prod_{j\in g(B)}r_j\bigg\rangle},
\end{equation}
for all $g\in S,A\subset\mathbb{Z}^2,B\subset\mathbb{Z}^2$.

The $\mathbb{Z}_2$ symmetry is on a different footing. A signature of ferromagnetism is the existence of a measure for which the expectation values of $\mathbb{Z}_2$-odd functions are nonzero. Therefore, when searching for such signatures, we should not impose $\mathbb{Z}_2$ symmetry. If instead we are interested in expectation values of $\mathbb{Z}_2$-even functions, we may impose $\mathbb{Z}_2$ symmetry in the form
\begin{equation}\label{eq:Z2symmetryAny}
    \overline{\bigg\langle w(r)\prod_{i\in A}s_i\bigg\rangle}=0~\text{if}~|A|~\text{is odd},
\end{equation}
for any functions $w(r)$. Choosing these functions to be indicator functions for $r_j=v_j$ over $j\in B\subset\mathbb{Z}^d$, we obtain
\begin{equation}\label{eq:Z2symmetry}
    \overline{\bigg\langle \prod_{j\in B}{1+v_jr_j\over2}\prod_{i\in A}s_i\bigg\rangle}=0~\text{if}~|A|~\text{is odd}.
\end{equation}

In the bootstrap formulation of the RSIM, which we detail later in (\ref{eq:RSIMLP}), disorder averages $\overline{\langle\cdots\rangle}$ are treated as bootstrap variables subject to the constraints discussed above. In practice, we impose only a finite subset of these constraints by restricting $A,B,W$ to a finite number of sets while considering all possible spin assignments $u_i$ and disorder assignments $v_j$ over them.

\subsection{Two-replica extension of RSIM}
One of the key observables in systems with quenched disorder is the replica correlator. In this work, we focus on two-replica correlators. At each site $i\in\mathbb{Z}^d$, we now have a pair of spins $s^{(1)}_i$ and $s^{(2)}_i$. The two-replica Hamiltonian is simply the sum of two individual Hamiltonians (\ref{eq:RSIMhamil}) that share the same disorder realization,
\begin{eqnarray}\label{eq:2RSIMhamil}
    {}&&H(s^{(1)},s^{(2)},r)=H^{(1)}+H^{(2)},~~\text{with}\nonumber
    \\
    &&H^{(m)}=-\sum_{(i,j)}n_in_js^{(m)}_is^{(m)}_j,~~m=1,2.
\end{eqnarray}
For a fixed disorder realization $r$, the two replicas are completely decoupled from each other. Therefore, for any function $f^{(1)}(s^{(1)})$ depending only on $s^{(1)}$ and any function $f^{(2)}(s^{(2)})$ depending only on $s^{(2)}$, the expectation value of their product factorizes as
\begin{equation}\label{eq:2factor}
    \langle f^{(1)}(s^{(1)})f^{(2)}(s^{(2)})\rangle_r=\langle f^{(1)}(s^{(1)})\rangle_r~\langle f^{(2)}(s^{(2)})\rangle_r.
\end{equation}
Therefore, for any function $f^{(1)}(s^{(1)},r)$ depending only on $s^{(1)}$ and $r$, and any function $f^{(2)}(s^{(2)},r)$ depending only on $s^{(2)}$ and $r$, we obtain
\begin{equation}\label{eq:2disorderAvg}
    \overline{\langle f^{(1)}(s^{(1)},r) f^{(2)}(s^{(2)},r)  \rangle}=\overline{\langle f^{(1)}(s^{(1)},r)\rangle_r~\langle f^{(2)}(s^{(2)},r)  \rangle}_r.
\end{equation}
However, because both replicas couple nontrivially to $r$, the disorder average (\ref{eq:2disorderAvg}) does not further factorize into a product of two disorder averages.

The positivity of the measure on the product space of the two replica spin configurations $s^{(1)}$ and $s^{(2)}$ and the site-dilution configuration $r$ is encoded by
\begin{equation}\label{eq:2PB}
    \overline{\langle I_{A^{(1)},A^{(2)},B;u^{(1)},u^{(2)},v}(s^{(1)},s^{(2)},r) \rangle}\geq0,
\end{equation}
where $I_{A^{(1)},A^{(2)},B;u^{(1)},u^{(2)},v}(s^{(1)},s^{(2)},r)$ is the indicator function for the spin assignments $s_{i^{(1)}}^{(1)}=u^{(1)}_{i^{(1)}}$ for ${i^{(1)}}\in A^{(1)}$ and $s_{i^{(2)}}^{(2)}=u^{(2)}_{i^{(2)}}$ for ${i^{(2)}}\in A^{(2)}$, together with the site-dilution assignments $r_j=v_j$ for $j\in B$. It is given by
\begin{eqnarray}\label{eq:2ind}
    {}&&I_{A^{(1)},A^{(2)},B;u^{(1)},u^{(2)},v}(s^{(1)},s^{(2)},r)\equiv \prod_{j\in B}{1+v_jr_j\over2}\nonumber
    \\
    &&~~~\times\prod_{{i^{(1)}}\in A^{(1)}}{1+u^{(1)}_{i^{(1)}}s^{(1)}_{i^{(1)}}\over2}\prod_{{i^{(2)}}\in A^{(2)}}{1+u^{(2)}_{i^{(2)}}s^{(2)}_{i^{(2)}}\over2}.~~~~~~~
\end{eqnarray}
The normalization condition is again given by (\ref{eq:norm}).

The disorder-averaged DLR equation (\ref{eq:DLRdis}) extends straightforwardly to the two-replica case as
\begin{widetext}
\begin{equation}\label{eq:2DLRdis1}
\overline{\bigg\langle I_{W\cup\{i\},A,B;u^{(1)},u^{(2)},v}(s^{(1)},s^{(2)},r)\bigg\rangle}=\overline{\bigg\langle I_{W,A,B;u^{(1)},u^{(2)},v}(s^{(1)},s^{(2)},r)\cdot~ \left[ 1+\exp\left(-2\beta r_{i}u^{(1)}_{i}\sum_{l\in N(i)}r_{l}u^{(1)}_{l}\right) \right]^{-1}\bigg\rangle},
\end{equation}
\begin{equation}\label{eq:2DLRdis2}
\overline{\bigg\langle I_{A,W\cup\{i\},B;u^{(1)},u^{(2)},v}(s^{(1)},s^{(2)},r)\bigg\rangle}=\overline{\bigg\langle I_{A,W,B;u^{(1)},u^{(2)},v}(s^{(1)},s^{(2)},r)\cdot~ \left[ 1+\exp\left(-2\beta r_{i}u^{(2)}_{i}\sum_{l\in N(i)}r_{l}u^{(2)}_{l}\right) \right]^{-1}\bigg\rangle},
\end{equation}
\end{widetext}
where $A$ is an arbitrary finite subset of $\mathbb{Z}^d$.

The disorder distribution (\ref{eq:disorderDistR}) remains unchanged, while the first Griffiths inequalities (\ref{eq:G1}) extend to
\begin{equation}\label{eq:2G1}
    \overline{\bigg\langle \left(\prod_{j\in B}{1+v_jr_j\over2}\right)\prod_{i^{(1)}\in A^{(1)}}s^{(1)}_{i^{(1)}}\prod_{i^{(2)}\in A^{(2)}}s^{(2)}_{i^{(2)}} \bigg\rangle}\geq0,
\end{equation}
for any subsets $A^{(1)}$ and $A^{(2)}$. For $d=2$, the spatial symmetry (\ref{eq:spatialsym}) extends to
\begin{eqnarray}\label{eq:2spatialsym}
    {}&&\overline{\bigg\langle\prod_{i^{(1)}\in A^{(1)}} s^{(1)}_{i^{(1)}}\prod_{i^{(2)}\in A^{(2)}} s^{(2)}_{i^{(2)}}\prod_{j\in B}r_j\bigg\rangle}\nonumber
    \\
    &&=\overline{\bigg\langle\prod_{i^{(1)}\in g(A^{(1)})} s^{(1)}_{i^{(1)}}\prod_{i^{(2)}\in g(A^{(2)})} s^{(2)}_{i^{(2)}}\prod_{j\in g(B)}r_j\bigg\rangle}.
\end{eqnarray}
The spin-flip $\mathbb{Z}_2$ symmetry (\ref{eq:Z2symmetry}) extends to a $\mathbb{Z}_2\times\mathbb{Z}_2$ symmetry:
\begin{equation}\label{eq:2Z2symmetry}
    \overline{\bigg\langle w(r)\prod_{i^{(1)}\in A^{(1)}} s^{(1)}_{i^{(1)}}\prod_{i^{(2)}\in A^{(2)}} s^{(2)}_{i^{(2)}}\bigg\rangle}=0,
\end{equation}
when $|A^{(1)}|$ or $|A^{(2)}|$ is odd. There is also a replica-exchange symmetry $S_2$ permuting the two replicas, leading to
\begin{eqnarray}\label{eq:2repSym}
    {}&&\overline{\bigg\langle w(r)\prod_{i^{(1)}\in A^{(1)}} s^{(1)}_{i^{(1)}}\prod_{i^{(2)}\in A^{(2)}} s^{(2)}_{i^{(2)}}\bigg\rangle}\nonumber
    \\
    &&=\overline{\bigg\langle w(r)\prod_{i^{(1)}\in A^{(1)}} s^{(2)}_{i^{(1)}}\prod_{i^{(2)}\in A^{(2)}} s^{(1)}_{i^{(2)}}\bigg\rangle}.
\end{eqnarray}
The spin-flip and replica-exchange symmetries combine into $(\mathbb{Z}_2\times\mathbb{Z}_2)\rtimes S_2$.

Finally, the factorization (\ref{eq:2disorderAvg}) provides a new type of positivity condition:
\begin{eqnarray}\label{eq:2squarepos}
    {}&&\overline{\langle z(r) f(s^{(1)})f(s^{(2)})\rangle}=\overline{z(r)\langle  f(s^{(1)})\rangle_r\langle f(s^{(2)})\rangle_r}\nonumber
    \\
    &&=\overline{z(r)\left(\langle f(s^{(1)})\rangle_r\right)^2}\geq0,
\end{eqnarray}
for any nonnegative function $z(r)$ of $r$ and any function $f(s)$. We may choose $z(r)$ to be an indicator function for the assignments $r_j=v_j$ over $j\in C\subset\mathbb{Z}^d$ and expand $f(s)$ in the spin-monomial basis. This yields PSD matrices whose matrix indices are labeled by subsets $A,B\subset\mathbb{Z}^d$:
\begin{eqnarray}\label{eq:2PSD}
    {}&&{\cal M}_{A,B}^{C;v}=\overline{\bigg\langle\left(\prod_{k\in C}{1+v_k r_k\over2}\right)\prod_{i\in A}s^{(1)}_i\prod_{j\in B}s^{(2)}_j\bigg\rangle}\nonumber
    \\
    &&~~~~\Rightarrow~{\cal M}^{C;v}\succeq0.
\end{eqnarray}
Therefore, there are infinitely many PSD matrices ${\cal M}^{C;v}$ labeled by subsets $C\subset\mathbb{Z}^d$ and site-dilution assignments $v$ on $C$, each of which is infinite-dimensional.

\subsection{Nonequilibrium example: DCP}
The contact process \cite{10.1214/aop/1176996493} is an interacting particle system \cite{liggett1985interacting} that models the spread of an epidemic. A site $i\in\mathbb{Z}^d$ is healthy if $s_i=-1$ and infected if $s_i=1$. As time evolves, the spin configuration $s$ changes randomly according to the following rule: if $s_i$ is infected, it recovers at rate 1, whereas if it is healthy, it becomes infected at rate $\lambda_i\times\text{(number of infected nearest neighbors)}$, where $\lambda_i$ is the infection rate. Given a spin configuration $s$, the transition rate $c_\lambda(i,s)$ for $s_i\rightarrow-s_i$ is therefore
\begin{equation}\label{eq:CPrate}
    c_\lambda(i,s)={1+s_i\over2}+\lambda_i{1-s_i\over2}\sum_{j\in N(i)}{1+s_j\over2}.
\end{equation}
The quenched disorder in the DCP \cite{10.1214/aop/1176990331} is parametrized by two parameters, $p\in[0,1]$ and $b\in[0,1]$. The variables $\lambda_i$ are i.i.d. across lattice sites and are drawn from $\{b\lambda,\lambda\}$, with $b\lambda$ chosen with probability $p$ and $\lambda$ chosen with probability $1-p$. In terms of $r_i$, $\lambda_i=\left({1+r_i\over2}+{1-r_i\over2}b\right)\lambda$, where $r_i=-1$ with probability $p$ and $r_i=1$ with probability $1-p$. With this convention, we denote the transition rate by $c_r(i,s)$ from here on.

For a fixed disorder realization $r$, expectation values evolve in time according to the Kolmogorov backward equation:
\begin{equation}\label{eq:CPmaster}
    {d\over dt}\langle f(s)\rangle_r=\sum_{i\in\mathbb{Z}^d}\bigg\langle c_r(i,s)\left(f(\tilde{s}^i)-f(s)\right)  \bigg\rangle_r,
\end{equation}
where $\tilde{s}^i$ is the same spin configuration as $s$, except that the spin at site $i$ is flipped, i.e., $\left(\tilde{s}^i\right)_j=(1-2\delta_{ij})s_j$. Therefore, the sum over $i\in\mathbb{Z}^d$ is finite as long as $f(s)$ depends on only finitely many spin variables. We focus on invariant measures whose expectation values satisfy the following invariance equations:
\begin{equation}\label{eq:CPrinv}
    \sum_{i\in\mathbb{Z}^d}\bigg\langle c_r(i,s)\left(f(\tilde{s}^i)-f(s)\right)  \bigg\rangle_r=0,
\end{equation}
for all functions $f(s)$ with finite support. These measures govern the late-time behavior of the system and are therefore of fundamental interest.

For any disorder realization $r$, there is a trivial invariant measure given by the so-called absorbing state, in which $s_i=-1,~\forall i\in\mathbb{Z}^d$. In the clean case $p=0$ (or $b=1$), there exists a critical point $\lambda=\lambda_0$ such that the absorbing state is the unique invariant measure for $\lambda\leq\lambda_0$, while nontrivial invariant measures appear for $\lambda>\lambda_0$. This absorbing phase transition exists even for $d=1$, although its exact solution remains unknown. For nontrivial values of $p$ and $b$, the infection rate is damped by a factor $b<1$ at randomly selected sites, and the critical value $\lambda_c$ of $\lambda$ increases. The region $\lambda_0\leq\lambda\leq\lambda_c$ provides another example of a Griffiths region. For example, for $p=0.3$ and $b=0.8$ in $d=1$, $\lambda_0\approx1.64892$ \cite{2005nptl.book.....M} and $\lambda_c\approx1.7625$ \cite{2005PhRvE..72c6126V}.

Following (\ref{eq:linearEq}), disorder-averaged invariant measures satisfy the following disorder-averaged invariance equations:
\begin{equation}\label{eq:CPdisInv}
    \sum_{i\in\mathbb{Z}^d}~\overline{\bigg\langle \left(\prod_{j\in B}{1+v_jr_j\over2}\right) c_r(i,s)\left(f(\tilde{s}^i)-f(s)\right)\bigg\rangle}=0,
\end{equation}
where $w(r)$ in (\ref{eq:linearEq}) is chosen to be an indicator function for $r_j=v_j$,~$j\in B\subset\mathbb{Z}^d$. When $B$ includes all sites $i$ for which $f(\tilde{s}^i)-f(s)\neq0$, the variable $r_i$ appearing in the transition rate $c_r(i,s)$ is replaced by $v_i$.

Information about the disorder distribution is encoded in the same way as in (\ref{eq:disorderDistR}). The variables $r_i$ are i.i.d. over $\mathbb{Z}^d$, giving
\begin{equation}\label{eq:CPdisorderDistr}
     \overline{\bigg\langle\prod_{j\in B}r_j\bigg\rangle}=(1-2p)^{|B|},~~\forall B\subset\mathbb{Z}^d.
\end{equation}
Note the difference in sign from (\ref{eq:disorderDistR}), which arises from our choice of convention.

For the contact process, the only symmetries are spatial symmetries. Focusing on the case $d=1$, the symmetry group $S$ is the semidirect product of the translation group $\mathbb Z$ and the reflection group $\mathbb{Z}_2$: $S=\mathbb{Z}\rtimes\mathbb{Z}_2$. The disorder averages of probability measures that respect $S$ satisfy
\begin{equation}\label{eq:spatialsym1d}
    \overline{\bigg\langle\prod_{i\in A} s_i\prod_{j\in B}r_j\bigg\rangle}=\overline{\bigg\langle\prod_{i\in g(A)} s_i\prod_{j\in g(B)}r_j\bigg\rangle},
\end{equation}
for all $g\in S,A\subset\mathbb{Z},B\subset\mathbb{Z}$.

\section{BOOTSTRAP SETUP AND RESULTS}
In the previous section, we reformulated the definitions and properties of the probability measures of interest in terms of their expectation values, i.e., disorder averages. In particular, all equations among these quantities were linear, while the inequalities were either linear or PSD constraints. They therefore provide convex constraints on the space of disorder averages defining systems with quenched disorder, leading to either LP or SDP formulations. In this section, we formulate the relevant LP and SDP problems for the systems of interest and present the resulting bootstrap bounds.

\textit{On the usage of generative AI.} OpenAI ChatGPT 5.5 and 5.6 were used as coding assistants for obtaining the bootstrap and Monte Carlo results reported below. For the bootstrap code, they assisted in extending the existing code for the clean case \cite{Cho:2022lcj,Cho:2025dgc} to the disordered case. For the Monte Carlo simulations, they generated initial versions of the code, which were subsequently reviewed by the authors. In both cases, the authors checked the final code line by line.

\subsection{RSIM}
We consider the RSIM on the $\mathbb{Z}^2$ lattice. In principle, there are infinitely many bootstrap constraints that could be imposed, so we need a reasonable truncation scheme in which only a finite subset of these constraints is included. Importantly, the bootstrap bounds obtained from such a truncation scheme remain rigorous and apply to a strictly infinite lattice, since we impose only constraints that must hold.

Suppose we are given a finite subset $D\subset\mathbb{Z}^2$ and a disorder average of interest $\overline{\langle {\cal O}(s,r) \rangle}$, where the support of ${\cal O}(s,r)$ lies within $D$, i.e., ${\cal O}(s,r)$ depends only on $s_i$ and $r_j$'s with $i,j\in D$. The LP problem \textbf{LP($D$)} that provides lower (upper) bootstrap bounds on $\overline{\langle {\cal O}(s,r) \rangle}$ is given by
\begin{eqnarray}\label{eq:RSIMLP}
    &&\text{\textbf{LP$(D)~$} minimize (maximize) }\overline{\langle {\cal O}(s,r) \rangle}\nonumber
    \\
    &&~~~\text{over $\overline{\bigg\langle\prod_{i\in A}s_i\prod_{j\in B}r_j\bigg\rangle}$ for all subsets $A,B\subset D$},\nonumber
    \\   
    &&~~~\text{subject to}\nonumber
    \\
    &&\text{Positivity: (\ref{eq:indPos}) with $A=B=D$},\nonumber
    \\
    &&\text{Normalization: (\ref{eq:norm})},\nonumber
    \\
    &&\text{DLR: (\ref{eq:DLRdisBW}) with $B=D$ for all $i\in D\setminus\partial D$}\nonumber
    \\
    &&~~~~~~~~~\text{and $W=D\setminus\{i\}$},\nonumber
    \\
    &&\text{Disorder: (\ref{eq:disorderDistR}) for all $B\subset D$},\nonumber
    \\
    &&\text{Griffiths: (\ref{eq:G1}) with $B=D$ for all $A\subset D$},\nonumber
    \\
    &&\text{Symmetry: (\ref{eq:spatialsym}) for all $A,B\subset D,~g\in S$},\nonumber
    \\
    &&\text{for all assignments $u,v$ appearing above.}
\end{eqnarray}
If we are interested in $\mathbb{Z}_2$ spin-flip-symmetric measures, we can further add the constraints (\ref{eq:Z2symmetry}) with $B=D$ for all assignments of $v$ over $B$ and all subsets $A\subset D$, and call the resulting LP problem \textbf{LP${}_{\mathbb{Z}^2}(D)$}. Note that the symmetry action may produce disorder averages of functions whose support lies outside the region $D$. We simply omit such constraints.

In practice, we used two choices of $D$ to obtain bootstrap bounds. These are the two diamond-shaped subsets $D_{131}$ and $D_{1331}$ of $\mathbb{Z}^2$, given by
\begin{eqnarray}
    {}&&D_{131}=\{(-1,0),(0,1),(0,0),(0,-1),(1,0)\},\nonumber
    \\
    &&D_{1331}=D_{131}\cup\{(1,1),(1,-1),(2,0)\}.\nonumber
\end{eqnarray}
One practical caveat is that the Griffiths inequalities quickly become computationally expensive as $|D|$ grows. Therefore, when solving \textbf{LP${}(D_{1331})$} and \textbf{LP${}_{\mathbb{Z}^2}(D_{1331})$}, we instead imposed (\ref{eq:G1}) with $B=D_{131}$ for all $A\subset D_{131}$.

The problems \textbf{LP($D$)} and \textbf{LP${}_{\mathbb{Z}^2}(D)$} can be solved using standard LP solvers. When all inputs are exact rational numbers, the \texttt{LinearOptimization} function in Mathematica \cite{Mathematica} provides exact rational solutions. For example, at $p={1\over5}$ and $e^\beta={3\over2}$, \textbf{LP${}_{\mathbb{Z}^2}(D_{131})$} produces the following bounds on the disorder average of the nearest-neighbor spin product $s_is_{i+e_1}$, where $e_1=(1,0)$ is the unit vector along the $x-$axis:
\begin{equation}
    {12419281\over859623700}\leq\overline{\langle s_is_{i+e_1} \rangle}\leq{56857093\over2621852285}.
\end{equation}
For the larger region $D_{1331}$, we instead used the double-precision MOSEK solver with its default parameters.

\begin{figure}
\centering
\includegraphics[width=8.6cm]{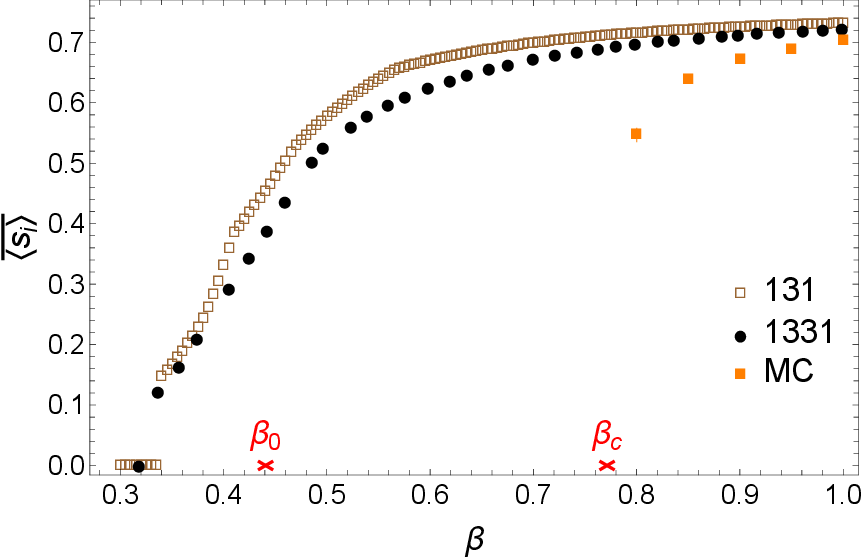}
\caption{\label{fig:RSIMmag} Upper bounds on $\overline{\langle s_i\rangle}$ at $p=0.75$ obtained from \textbf{LP${}(D_{131})$} (brown) and \textbf{LP${}(D_{1331})$} (black), together with Monte Carlo estimates (orange) obtained using the Metropolis algorithm for 200 disorder realizations on a $50\times50$ periodic lattice with 100000 sweeps. Also shown are the critical values of $\beta$ for the clean case $(\beta_0={\log(1+\sqrt{2})\over2})$ and for $p=0.75$ ($\beta_c\approx0.771356$) \cite{2026arXiv260321303B}.}
\end{figure}

In Fig. \ref{fig:RSIMmag}, we present bootstrap upper bounds on the magnetization $\overline{\langle s_i\rangle}$ obtained from \textbf{LP($D_{131}$)} and \textbf{LP($D_{1331}$)}, together with Monte Carlo estimates. The lower bounds are simply the negatives of the upper bounds. The upper bounds are identically zero for $\beta$ up to ${\log(1+2\sqrt{2})\over4} \approx 0.336$, as in the clean case, for which this result was shown analytically \cite{Cho:2022lcj}. Therefore, the current bootstrap bounds on the disorder average do not appear to provide a nontrivial lower bound on the critical value $\beta_c$ that is beyond the clean case since ${\log(1+2\sqrt{2})\over4}<\beta_0<\beta_c$. Moreover, there is a wide range of $\beta$ over which the bootstrap bounds remain far from the expected values. This is not surprising since bootstrap results obtained from finite small regions like $D_{1331}$ are not expected to have access to full information about the system especially near Griffiths region and criticality. The mild improvement of bounds from \textbf{LP($D_{131}$)} to \textbf{LP($D_{1331}$)} around this region is expected to be a reflection of the physics of the Griffiths region where the equilibration time of rare regions is exponentially long. Nonetheless, these bounds are nontrivial for $\beta>\beta_c$, where they approach the Monte Carlo estimates as $\beta$ increases. One curious feature of these upper bounds is the presence of multiple kinks above $\beta={\log(1+2\sqrt{2})\over4}$, which were absent from the clean-case bounds in \cite{Cho:2022lcj}. We will observe a similar feature later for the DCP, which we will relate to the Griffiths region.

\begin{figure}
\centering
\includegraphics[width=8.6cm]{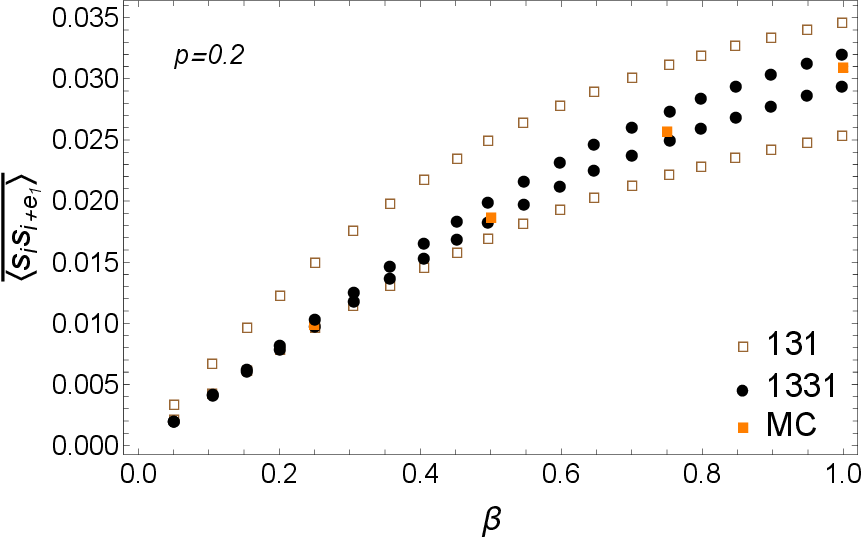}
\\
\includegraphics[width=8.6cm]{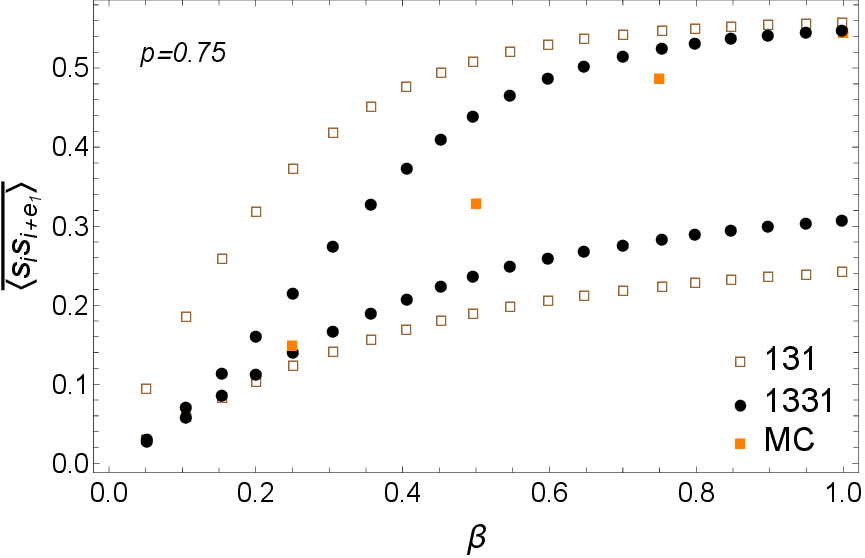}
\caption{\label{fig:RSIMnn} Lower and upper bounds on $\overline{\langle s_is_{i+e_1}\rangle}$ at $p=0.2$ (top) and $p=0.75$ (bottom) obtained from \textbf{LP${}_{\mathbb{Z}_2}(D_{131})$} (brown) and \textbf{LP${}_{\mathbb{Z}_2}(D_{1331})$} (black), together with Monte Carlo estimates (orange) obtained using the Metropolis algorithm for 200 disorder realizations on a $50\times50$ periodic lattice with 100000 sweeps.}
\end{figure}

In Fig. \ref{fig:RSIMnn}, we further present bootstrap bounds on the nearest-neighbor correlator $\overline{\langle s_is_{i+e_1}\rangle}$ obtained from \textbf{LP${}_{\mathbb{Z}_2}(D_{131})$} and \textbf{LP${}_{\mathbb{Z}_2}(D_{1331})$}. Even in a regime where the ferromagnetic phase is absent, e.g., $p=0.2$, this $\mathbb{Z}_2$-invariant observable remains nontrivial. We observe that the bounds are much tighter for smaller values of $p$, i.e., stronger site dilution.

We remark that among the constraints in (\ref{eq:RSIMLP}), the Griffiths inequalities are distinct from the others because they are not part of the definition of the Gibbs measures of interest. Therefore, they are not expected to be essential for obtaining tight bootstrap bounds once enough number of other bootstrap constraints are imposed. When only a finite subset of bootstrap constraints are imposed though, the Griffiths inequalities may still lead to nontrivial improvements of the bounds. In Appendix \ref{sec:G1}, we study the effect of the Griffiths inequalities on bootstrap bounds by comparing bounds obtained with and without the first Griffiths inequalities.

\subsection{Two-replica RSIM}
\begin{figure}
\centering
\includegraphics[width=8.6cm]{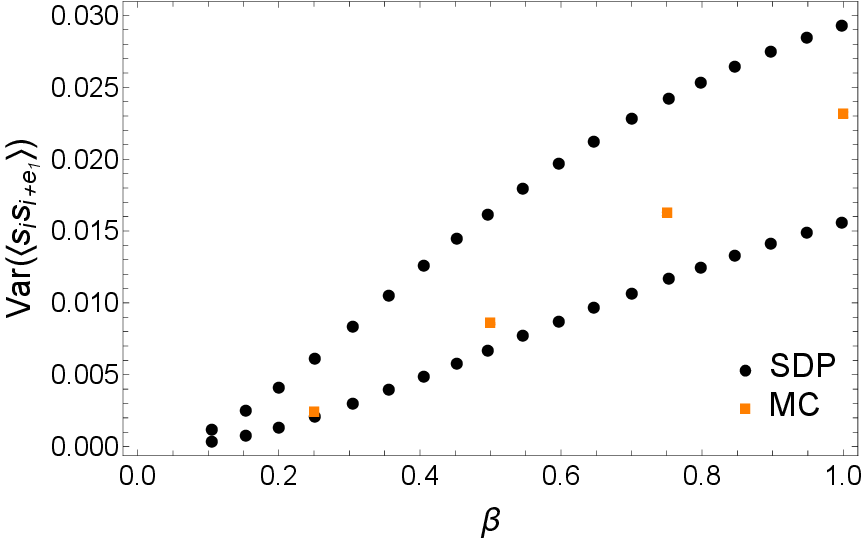}
\caption{\label{fig:2RSIM} Lower and upper bounds (black) on $\text{Var}(\langle s_is_{i+e_1}\rangle)$ at $p=0.2$ obtained from \textbf{SDP${}_{\mathbb{Z}_2}(D_{131})$}, together with Monte Carlo estimates (orange) obtained using the Metropolis algorithm for 200 disorder realizations on a $50\times50$ periodic lattice with 30000 sweeps.}
\end{figure}
Bootstrapping the two-replica RSIM proceeds similarly to the single-replica case (\ref{eq:RSIMLP}), except that the two-replica Gram matrices (\ref{eq:2PSD}) provide additional constraints, leading to an SDP rather than an LP. Given a finite subset $D\subset\mathbb{Z}^2$ and a disorder average of interest $\overline{\langle {\cal O}(s^{(1)},s^{(2)},r) \rangle}$ whose support lies within $D$, we define \textbf{SDP$(D)$} by the following SDP problem:
\begin{eqnarray}\label{eq:2RSIMSDP}
    &&\text{\textbf{SDP$(D)~$} minimize (maximize) }\overline{\langle {\cal O}(s^{(1)},s^{(2)},r) \rangle}\nonumber
    \\
    &&~~~\text{over $\overline{\bigg\langle\prod_{i^{(1)}\in A^{(1)}}s_{i^{(1)}}\prod_{i^{(2)}\in A^{(2)}}s_{i^{(2)}}\prod_{j\in B}r_j\bigg\rangle}$}\nonumber
    \\
    &&~~~\text{for all subsets $A^{(1)},A^{(2)},B\subset D$},\nonumber
    \\
    &&~~~\text{subject to}\nonumber
    \\
    &&\text{Positivity: (\ref{eq:2PB}) with $A^{(1)}=A^{(2)}=B=D$},\nonumber
    \\
    &&\text{Normalization: (\ref{eq:norm})},\nonumber
    \\
    &&\text{DLR: (\ref{eq:2DLRdis1}) and (\ref{eq:2DLRdis2}) with $A=B=D$}\nonumber
    \\
    &&~~~~~~~~~\text{for all $i\in D\setminus\partial D$ and $W=D\setminus\{i\}$},\nonumber
    \\
    &&\text{Disorder: (\ref{eq:disorderDistR}) for all $B\subset D$},\nonumber
    \\
    &&\text{Griffiths: (\ref{eq:2G1}) with $B=D$ for all $A^{(1)},A^{(2)}\subset D$},\nonumber
    \\
    &&\text{Symmetry: (\ref{eq:2spatialsym}) for all $A^{(1)},A^{(2)},B\subset D,~g\in S$},\nonumber
    \\
    &&\text{$S_2$:~(\ref{eq:2repSym}) with $w(r)=\prod_{j\in D}{1+v_jr_j\over2}$ for all $A^{(1)},A^{(2)}\subset D$},\nonumber
    \\
    &&\text{PSD: (\ref{eq:2squarepos}) for all $C\subset D$,}\nonumber
    \\
    &&~~~~~~~~\text{with $A,B$ ranging over all subsets of $D$},\nonumber
    \\
    &&\text{for all assignments $u,v$ appearing above.}
\end{eqnarray}
If we are interested in a measure that respects the $\mathbb{Z}_2\times \mathbb{Z}_2$ spin-flip symmetry, we further impose (\ref{eq:2Z2symmetry}) with $w(r)=\prod_{j\in D}{1+v_jr_j\over2}$ for all $v_j$ assignments over $D$ and all $A^{(1)},A^{(2)}\subset D$ such that $|A^{(1)}|$ or $|A^{(2)}|$ is odd. The resulting SDP is denoted by \textbf{SDP${}_{\mathbb{Z}_2}(D)~$}. We obtained the corresponding bootstrap bounds on the two-replica nearest-neighbor correlator $\overline{\langle s_i^{(1)}s_{i+e_1}^{(1)}s_i^{(2)}s_{i+e_1}^{(2)}\rangle}=\overline{\langle s_i^{(1)}s_{i+e_1}^{(1)}\rangle^2}$ at $p=0.2$ for $D=D_{131}$ using MOSEK. Combining these bounds with the \textbf{LP${}_{\mathbb{Z}_2}(D_{1331})$} bounds on $\overline{\langle s_is_{i+e_1}\rangle}$ in Fig. \ref{fig:RSIMnn} yields bootstrap bounds on the disorder variance of the nearest-neighbor correlator $\text{Var}(\langle s_is_{i+e_1}\rangle)=\overline{\langle s_is_{i+e_1}\rangle^2}-\left(\overline{\langle s_is_{i+e_1}\rangle}\right)^2$, as presented in Fig. \ref{fig:2RSIM}.

\subsection{DCP}
\begin{figure}
\centering
\includegraphics[width=8.6cm]{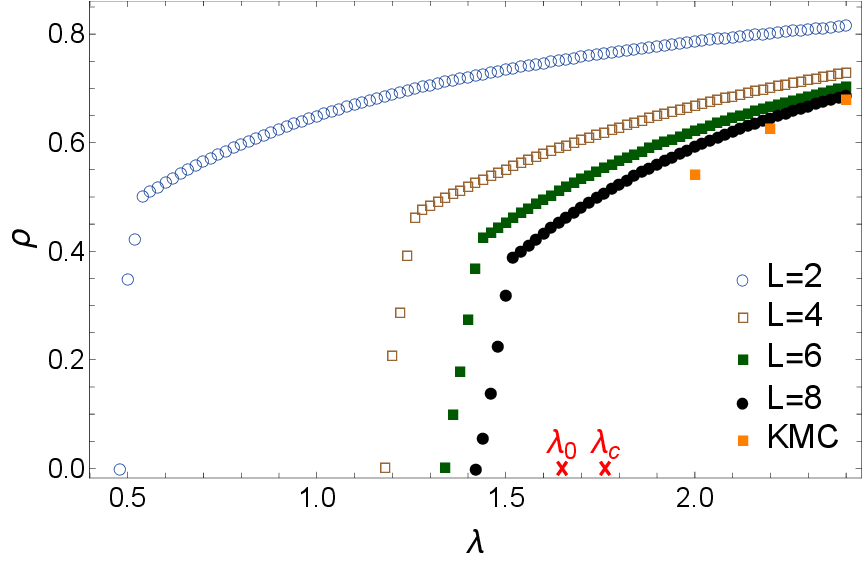}
\caption{\label{fig:CPrho} Upper bounds on the infection density $\rho=\overline{\bigg\langle {1+s_i\over2}\bigg\rangle}$ at $p=0.3$ and $b=0.8$ obtained from \textbf{LP${}_{DCP}(L)$}, together with KMC estimates (orange) obtained from 200 disorder realizations and 20 independent runs per disorder realization on a periodic lattice with 200 sites and a maximum simulation time of $t=4000$. Also shown are the critical values of $\lambda$ for the clean case $(\lambda_0\approx1.64892)$ \cite{2005nptl.book.....M} and for the present disordered case $(\lambda_c\approx1.7625)$ \cite{2005PhRvE..72c6126V}.}
\end{figure}
We consider the DCP on the $\mathbb Z$ lattice. To truncate the bootstrap constraints systematically, we introduce the notation
\begin{equation}
    D^{(L)}=\{0,1,2,3,\cdots,L\}\subset\mathbb Z.
\end{equation}
We consider bootstrap constraints that close within $D^{(L)}$. Given a disorder average of interest $\overline{\langle {\cal O}(s,r) \rangle}$ whose support lies within $D^{(L-2)}$, we define \textbf{LP${}_{DCP}(L)$} by the following LP problem:
\begin{eqnarray}\label{eq:DCPLP}
    &&\text{\textbf{LP${}_{DCP}(L)~$} minimize (maximize) }\overline{\langle {\cal O}(s,r) \rangle}\nonumber
    \\
    &&~~~\text{over $\overline{\bigg\langle\prod_{i\in A}s_i\prod_{j\in B}r_j\bigg\rangle}$ for all subsets $A\subset D^{(L-1)}$}\nonumber
    \\
    &&~~~\text{and $B\subset D^{(L-2)}$, subject to}\nonumber
    \\
    &&\text{Positivity: (\ref{eq:indPos}) with $A=D^{(L-1)},B=D^{(L-2)}$},\nonumber
    \\
    &&\text{Normalization: (\ref{eq:norm})},\nonumber
    \\
    &&\text{Invariance: (\ref{eq:CPdisInv}) with $B=D^{(L-2)}$}\nonumber
    \\
    &&~~~~~~~~~~~~~~~~\text{and $f(s)=\prod_{k\in A}s_k$ for all $A\subset D^{(L-2)}$},\nonumber
    \\
    &&\text{Disorder: (\ref{eq:CPdisorderDistr}) for all $B\subset D^{(L-2)}$},\nonumber
    \\
    &&\text{Symmetry:~(\ref{eq:spatialsym1d}) for all $A\subset D^{(L-1)},B\subset D^{(L-2)},g\in S$},\nonumber
    \\
    &&\text{for all assignments $u,v$ appearing above.}
\end{eqnarray}
The $s$ configuration is over $D^{(L-1)}$, whereas the $r$ configuration is over $D^{(L-2)}$, because the transition rate $c_r(i,s)$ appearing in the invariance equations depends on the nearest-neighbor spins of $i$ but not on the nearest-neighbor infection rates. In addition, the invariance equations produce a disorder average of a function involving the spin at site $-1\in\mathbb{Z}$. We shift this site to the origin using translation invariance so that the resulting disorder average remains within the variable space of the LP. Apart from this case, we ignore constraints obtained from symmetry actions that produce disorder averages lying outside the variable space.

\begin{figure}
\centering
\includegraphics[width=8.6cm]{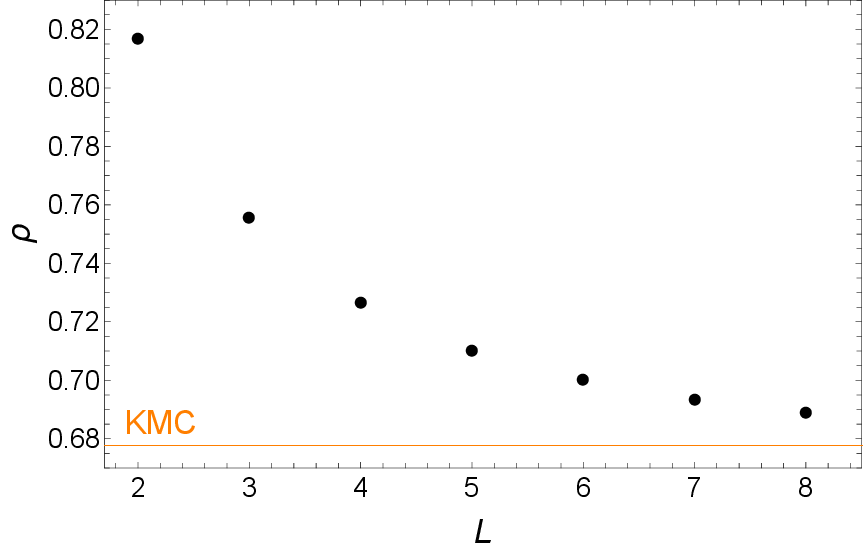}
\caption{\label{fig:CPfixLam} Upper bounds (black) on the infection density $\rho=\overline{\bigg\langle {1+s_i\over2}\bigg\rangle}$ at $p=0.3$, $b=0.8$, and $\lambda=2.4$ obtained from \textbf{LP${}_{DCP}(L)$}, shown as a function of $L$, together with the KMC estimate (orange line).}
\end{figure}

In Fig. \ref{fig:CPrho}, we present upper bounds on the infection density $\rho=\overline{\bigg\langle {1+s_i\over2}\bigg\rangle}$ at $p=0.3$ and $b=0.8$, obtained from \textbf{LP${}_{DCP}(L)$} using MOSEK. Note that the lower bounds are identically zero because of the presence of the absorbing state. In the supercritical regime $\lambda>\lambda_c$, the bootstrap upper bounds are genuinely nontrivial and approach the kinetic Monte Carlo (KMC) estimates as $L$ increases, as shown in greater detail in Fig. \ref{fig:CPfixLam} for $\lambda=2.4$.

\begin{figure}
\centering
\includegraphics[width=8.6cm]{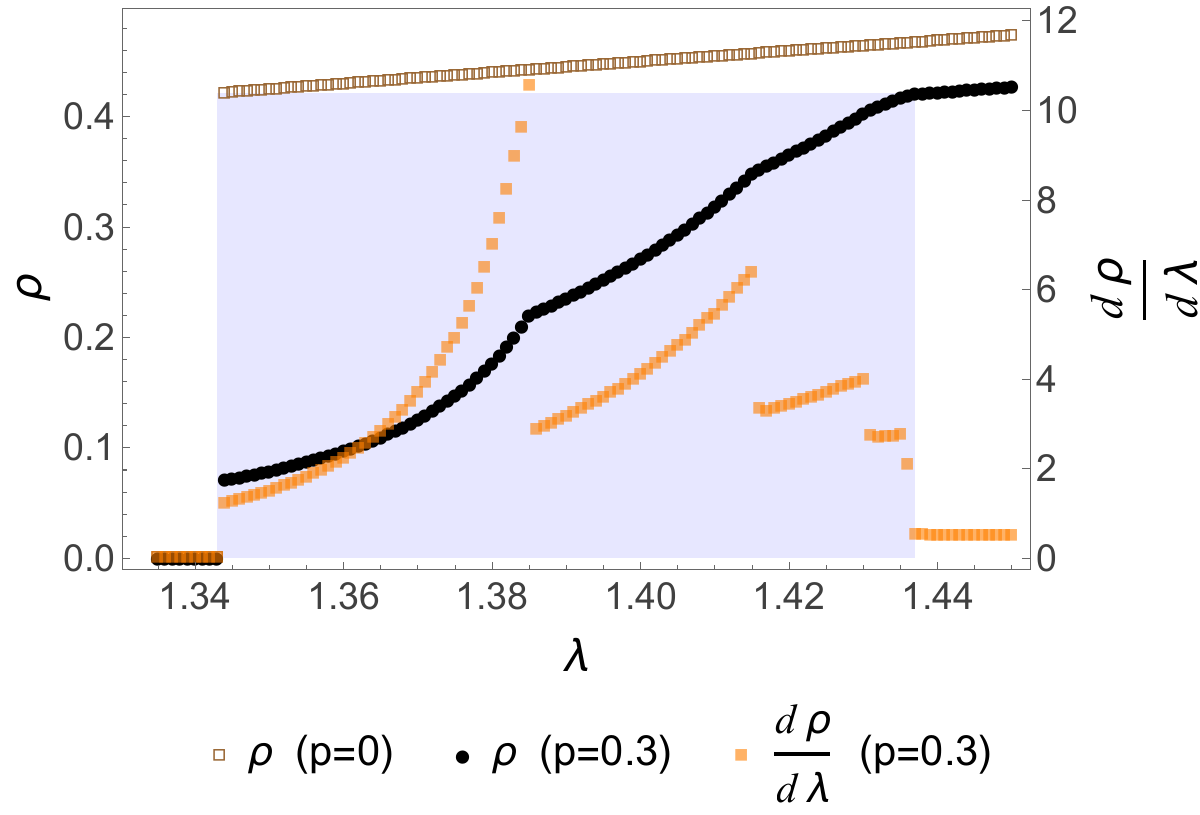}
\caption{\label{fig:CPkink} Close-up of the \textbf{LP${}_{DCP}(6)$} upper bounds on the infection density $\rho=\overline{\bigg\langle {1+s_i\over2}\bigg\rangle}$ at $b=0.8$ for the clean case $p=0$ (brown) and the disordered case $p=0.3$ (black). Also shown are the derivatives ${d\rho\over d\lambda}$ of the \textbf{LP${}_{DCP}(6)$} upper bounds on $\rho$ with respect to $\lambda$ at $p=0.3$ (orange), obtained using the envelope theorem. Within the shaded blue region, the bounds for $p=0.3$ exhibit multiple kinks as $\lambda$ increases before eventually entering a smooth, kink-free region.}
\end{figure}

When $\lambda$ is sufficiently small, the bootstrap upper bounds on $\rho$ are zero within the precision of the solver, thereby producing lower bounds on the critical value $\lambda_c$. However, at each $L$, we find that these lower bounds are identical to the lower bounds on $\lambda_0$ for the clean case obtained in \cite{Cho:2025dgc}. This suggests that the current bootstrap for disorder averages does not provide lower bounds on the critical value $\lambda_c$ beyond those already obtained for the clean case, similarly to the RSIM case. This can be similarly attributed to the difficulty of obtaining tight bounds in the Griffiths region using a bootstrap region with finite support.

Instead, there is a curious range of $\lambda$ in which the upper bounds exhibit multiple kinks that were absent in the clean case of $p=0$ studied in \cite{Cho:2025dgc}. In Fig. \ref{fig:CPkink}, we compare the bounds for $p=0$ and $p=0.3$ at $L=6$. In contrast to the bounds for $p=0$, which remain smooth after passing the discontinuity following the $\rho=0$ region, the bounds for $p=0.3$ exhibit multiple kinks before reaching the smooth region. The range of $\lambda$ between the first and last kinks encountered as $\lambda$ increases is colored blue, and we refer to it as the kink region. The figure also shows the derivative of the upper bounds on $\rho$ for $p=0.3$ with respect to $\lambda$, which helps diagnose the kink region. This derivative can be conveniently obtained from the primal and dual solutions of \textbf{LP${}_{DCP}(6)$} using the envelope theorem (see Appendix \ref{sec:envelope} for details).

Kinks observed in bootstrap bounds as external parameters such as $\lambda$ change are signatures of changes in the active constraints. In optimization problems such as \textbf{LP${}_{DCP}(L)$}, active constraints are inequality constraints that are saturated at the optimal solution. Since the inequality constraints of \textbf{LP${}_{DCP}(L)$} are positivity constraints on disorder averages of indicator functions, such active constraints correspond to zero-probability events. For example, $\rho=\overline{\bigg\langle {1+s_i\over2}\bigg\rangle}$ is the probability that $s_i=1$. Its bootstrap upper bound is identically zero for sufficiently small $\lambda$, where $\overline{\bigg\langle {1+s_i\over2}\bigg\rangle}\geq0$ is active, but becomes nonzero beyond the first kink, where $\overline{\bigg\langle {1+s_i\over2}\bigg\rangle}\geq0$ becomes inactive. Therefore, the presence of multiple kinks indicates that certain probabilities transition from zero to nonzero within the kink region.

On the strictly infinite lattice, $\rho$ of the invariant measure is identically zero for $\lambda\leq\lambda_c$, which includes the kink region. However, \textbf{LP${}_{DCP}(L)$} at a finite value of $L$ may allow for a probability measure in which $\rho$ or other probabilities involving some spins taking values $+1$ are nonzero even for $\lambda\leq\lambda_c$. At the lower end of the kink region, $\rho$ is allowed to be nonzero. However, the probability that multiple spins take values $+1$ may still be zero near the lower end if some of those spins are conditioned to carry the damped infection rate $b\lambda$, as allowed by the quenched disorder in DCP. As $\lambda$ increases, however, such probabilities may eventually transition from zero to nonzero within the kink region as the increase in $\lambda$ overcomes the infection damping by $b<1$. This provides a mechanism for the presence of the multiple kinks observed in Fig. \ref{fig:CPkink}.

\begin{figure}
\centering
\includegraphics[width=8.6cm]{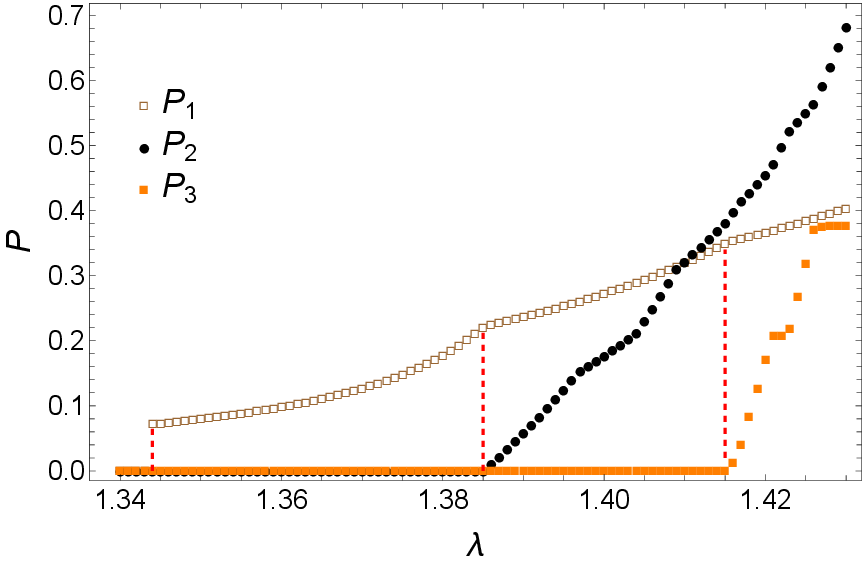}
\caption{\label{fig:CPprobs} Probabilities evaluated at the optimal solution obtained by maximizing $\rho$ in \textbf{LP${}_{DCP}(6)$}: $P_1=\rho=P(s_i=1)$ (brown), $P_2=P(s_0+s_1+s_2+s_3\neq-4~|~r_0+r_1+r_2=-1)$ (black), and $P_3=P(s_0+s_1+s_2+s_3\neq-4~|~r_0=r_1=r_2=-1)$ (orange). The dotted red lines indicate the values of $\lambda$ at which kinks appear in Fig. \ref{fig:CPkink}.}
\end{figure}

By evaluating the probabilities, which are disorder averages of indicator functions, at the optimal solutions obtained by maximizing $\rho$ in \textbf{LP${}_{DCP}(6)$}, we observe that many probabilities change from zero (active) to nonzero (inactive) across the kinks. In Fig. \ref{fig:CPprobs}, we show two such examples together with $P_1=\rho=P(s_i=1)$, which was already presented in Fig. \ref{fig:CPkink}. The probability $P_2=P(s_0+s_1+s_2+s_3\neq-4~|~r_0+r_1+r_2=-1)$ that at least one of $\{s_0,s_1,s_2,s_3\}$ takes the value $+1$, conditioned on exactly two of $\{r_0,r_1,r_2\}$ being $-1$, i.e., with two sites damped, changes from zero to nonzero as $\lambda$ crosses the kink at $\lambda\approx1.385$. Similarly, the probability $P_3=P(s_0+s_1+s_2+s_3\neq-4~|~r_0=r_1=r_2=-1)$ that at least one of $\{s_0,s_1,s_2,s_3\}$ takes the value $+1$, conditioned on all of $\{r_0,r_1,r_2\}$ being $-1$, i.e., with three sites damped, changes from zero to nonzero as $\lambda$ crosses the kink at $\lambda\approx1.415$. In contrast, for the clean case, the probability that at least one of $\{s_0,s_1,s_2,s_3\}$ takes the value $+1$ is nonzero throughout the entire kink region.

In fact, these observations about the kink region are reminiscent of the Griffiths region. For the physical invariant measure of the DCP on the strictly infinite lattice, the probability that at least one spin takes the value $+1$ is nonzero for the clean case ($b=1$), while it is zero for the disordered case ($b<1$), when $\lambda$ lies within the Griffiths region. However, confirming the latter for the invariant measure of the DCP, for example via Monte Carlo simulations, requires running the simulations over a time scale that is exponentially large in the size of the rare active spatial regions allowed by the quenched disorder, implying that such a probability can remain nonzero for a long simulation time. An analogous feature of the bootstrap problem \textbf{LP${}_{DCP}(L)$} at a fixed finite value $L$ is that there exist probability measures satisfying all the bootstrap constraints while giving a nonzero value for such a probability in the kink region.

\begin{figure}
\centering
\includegraphics[width=8.6cm]{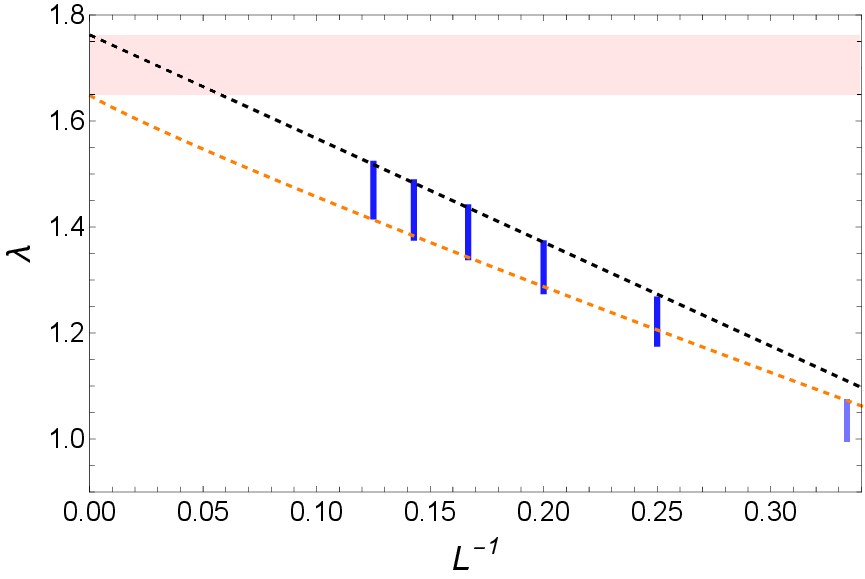}
\caption{\label{fig:CPkinkRegion} Kink regions for different values of $L$ (blue), obtained by scanning over $\lambda$ with grid spacing $0.001$, together with the Griffiths region (red), $\lambda\in[\lambda_0,\lambda_c].$ Also shown are the fit $\lambda(L)=\lambda_c+{a_c\over L}$ (dotted black), obtained from the last kinks at $L=5,\cdots,8$ with $\lambda_c=1.7625$, and the fit $\lambda(L)=\lambda_0+{a_0\over L^{0.912}}$ (dotted orange), obtained from the tightest lower bounds on $\lambda_0$ at $L=5,\cdots,10$ computed in \cite{Cho:2025dgc} with $\lambda_0=0.164892$.}
\end{figure}

\begin{table}[t]
\centering
\begin{tabular}{c|c|c}
\hline\hline
\diagbox{Fit range}{Parameters} & $p=0.3,b=0.8$ & $p=0.3, b=0.6$ \\
\hline
$2\leq L\leq5$ & 1.295 & 1.309\\
$3\leq L\leq6$ & 1.099 & 1.127\\
$4\leq L\leq7$ & 1.049 & 1.080\\
$5\leq L\leq8$ & 1.023 & 1.056\\
\hline
Expected & \multicolumn{2}{c}{$1/\tilde\nu=1$}\\
\hline\hline
\end{tabular}
\caption{Values of the exponent $\alpha$ obtained by fitting the upper ends of the kink region to $\lambda(L)=\lambda_c+{a_c\over L^\alpha}$ with different fit ranges in $L$, for two sets of disorder parameters, $(p,b)=(0.3,0.8)$ and $(0.3,0.6)$, with $\lambda_c=1.7625$ and $1.915$, respectively \cite{2005PhRvE..72c6126V}.}
\label{tab:alpha_fits}
\end{table}

A natural expectation is that the kink region approaches the Griffiths region $[\lambda_0,\lambda_c]$ as $L$ increases, with the rate of convergence governed by the critical exponents. For the clean case, we expect the lower ends of the kink region to approach $\lambda_0$ as $\lambda(L)=\lambda_0+{a_0\over L^{1/\nu_{\perp}}}$ at large $L$, with nonuniversal $a_0$, where $\nu_{\perp}\approx1.097$ is the spatial correlation length exponent \cite{2000AdPhy..49..815H}. For the infinite-randomness fixed point of the disordered case, we expect the upper ends of the kink region to approach $\lambda_c$ as $\lambda(L)=\lambda_c+{a_c\over L}$ at large $L$, with nonuniversal $a_c$, where the precise power ${1\over L}$ is determined by the typical correlation length exponent ${\tilde\nu}=1$ \cite{PhysRevLett.69.534}. Figure \ref{fig:CPkinkRegion} shows the kink regions for different values of $L$, together with the expected behaviors of $\lambda(L)$ obtained by fitting the $L=5,\cdots,10$ results for the lower ends of the kink region (from the bounds presented in \cite{Cho:2025dgc}) and the $L=5,\cdots,8$ results for the upper ends of the kink region, with $a_0$ and $a_c$ as the respective fitting parameters. These results suggest that the kink regions indeed approach the Griffiths region with the expected critical exponents.

Compared with the lower ends, which have not yet fully settled into the expected behavior at the displayed values of $L$, the upper ends quickly approach the expected curve as $L$ increases. In Table \ref{tab:alpha_fits}, we extract the typical correlation length exponent $\tilde\nu$ by performing a finite-size scaling analysis in which we fit the upper ends of the kink regions as $\lambda(L)=\lambda_c+{a_c\over L^{\alpha}}$, with $a_c$ and $\alpha$ as fitting parameters, using $L=k,\cdots,k+3$ for $k$ ranging from 2 to 5, for two sets of disorder parameters, $(p,b)=(0.3,0.8)$ and $(0.3,0.6)$. In both cases, the resulting values of $\alpha$ indeed approach the expected value $1/\tilde\nu=1$ as $k$ increases.

\section{Discussion}
In this work, we introduced a bootstrap framework for classical statistical systems with quenched disorder and applied it to the RSIM and the DCP. Although these bounds are not always tight, they nonetheless provide nontrivial bounds on disorder averages for the infinite lattice. These bounds are also mathematically rigorous when the LP is implemented using exact rational numbers. For the DCP example, we argued that the region of multiple kinks approaches the Griffiths region with the expected critical exponents, as supported by finite-size scaling analysis. We conclude by discussing several future directions for this framework.

The main computational bottleneck is the exponentially growing size of the bootstrap problem. At each lattice site $i$, we have not only the spin $s_i$ but also the disorder variable $r_i$. Therefore, bootstrap constraints confined to a region with $L$ sites involve $\sim 4^L$ variables, and the scaling becomes even worse when multiple replicas are included. It will be important to identify which of these exponentially many bootstrap constraints are the most physically relevant. In recent years, there have been several encouraging results on the quantum spin-chain bootstrap \cite{Kull:2022wof,Cho:2024owx}, in which tensor networks have been used to identify physically relevant bootstrap variables and constraints that grow only linearly, producing stronger bounds. Similar strategies for classical statistical systems will be necessary to obtain substantially stronger bounds than those reported in the present work.

An intriguing class of systems with quenched disorder consists of frustrated systems with disorder in which the couplings are not ferromagnetic but instead take both signs, such as spin glasses \cite{SFEdwards_1975,PhysRevLett.35.1792,Mezard1987}. Obtaining rigorous or precise results for such systems has been notoriously difficult. The bootstrap framework introduced here can be applied straightforwardly to these systems and can, in principle, provide rigorous results, even if the resulting bounds may not be as strong as desired.

Finally, quantum systems with quenched disorder also admit a similar bootstrap formulation as we report in \cite{unpubl1}, where we extend the ordinary quantum mechanical bootstrap for clean systems \cite{PhysRevA.57.4219,10.1063/1.1360199,Barthel:2012mqo,Han:2020bkb,Nancarrow:2022wdr,Fawzi:2023fpg,Gao:2024etm,Cho:2024kxn} to disordered systems.

\begin{acknowledgments}
We thank Akshat Panday, Pavel A. Nosov, and Zhaoyu Han for valuable conversations. The work of MC is supported by the Simons Collaboration on Global Categorical Symmetries. MGS is supported by a postdoctoral fellowship from the Julian Schwinger Foundation. Research reported in this publication was supported by an award from the Harvard FAS Dean’s Competitive Fund for Promising Scholarship. This work was completed in part with resources provided by the University of Chicago’s Research Computing Center. 
\end{acknowledgments}

\appendix

\section{Role of the first Griffiths inequalities}\label{sec:G1}
\begin{figure}
\centering
\includegraphics[width=8.6cm]{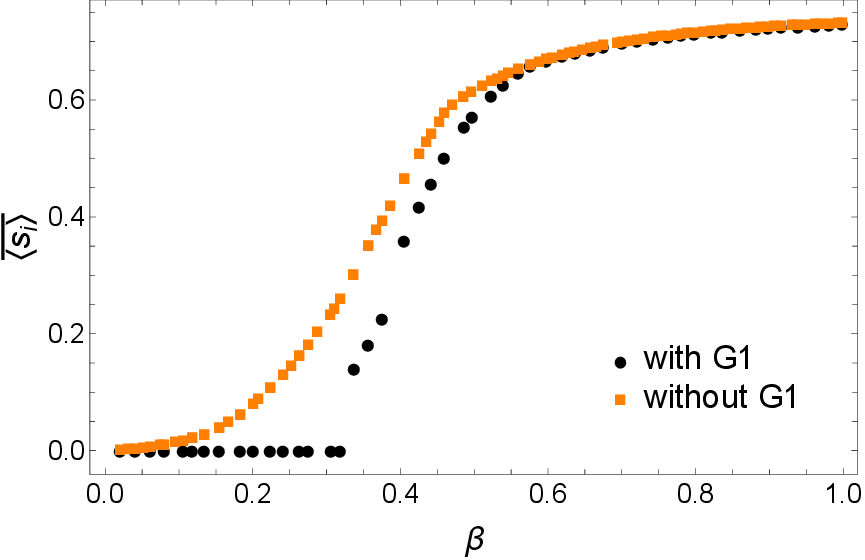}
\\
\includegraphics[width=8.6cm]{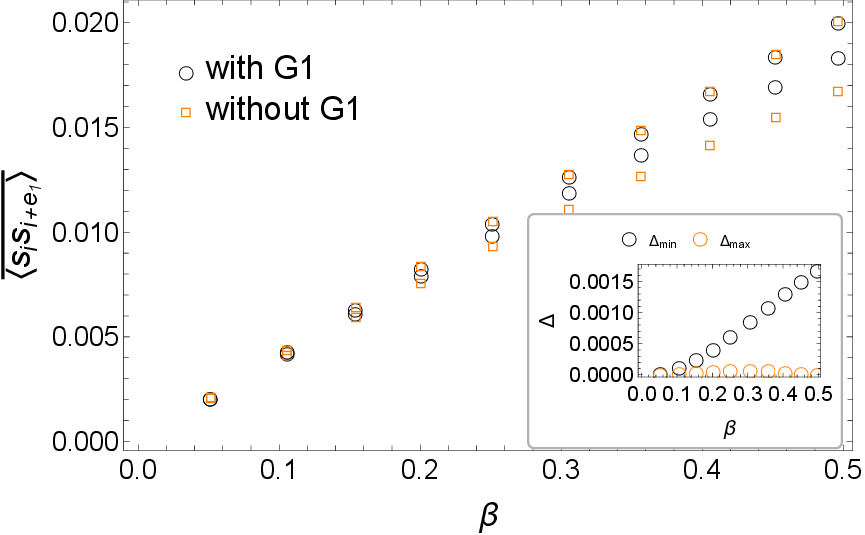}
\caption{\label{fig:G1} Top: Upper bounds on $\overline{\langle s_i\rangle}$ at $p=0.75$ obtained from \textbf{LP${}(D_{131})$} with (black) and without (orange) G1. Bottom: Lower and upper bounds on $\overline{\langle s_is_{i+e_1}\rangle}$ at $p=0.2$ obtained from \textbf{LP${}_{\mathbb{Z}_2}(D_{1331})$} with (black) and without (orange) G1 imposed over the $D_{131}$ region. The inset shows the differences $\Delta_{\text{min}}$ between the lower bounds obtained with and without G1 (black), and the differences $\Delta_{\text{max}}$ between the upper bounds obtained with and without G1 (orange).}
\end{figure}

Griffiths inequalities are not part of the definition of Gibbs measures but are instead derived properties. Therefore, they are not expected to be necessary for the convergence of the bootstrap bounds as the sublattice over which the bootstrap constraints are imposed grows, although they may help improve the bounds at a fixed sublattice size. In fact, the proof of convergence of the bootstrap bounds \cite{Cho:2023ulr} does not rely on Griffiths inequalities at all.

In Fig. \ref{fig:G1}, we present bootstrap bounds for the RSIM with and without the first Griffiths inequalities (G1). The top figure shows \textbf{LP${}(D_{131})$} upper bounds on $\overline{\langle s_i\rangle}$ at $p=0.75$, where G1 leads to a nontrivial lower bound on the critical value of $\beta$. For $\beta\gtrsim0.575$, however, the bounds with and without G1 become identical. Given that the infinite-randomness fixed point occurs at $\beta\approx0.771356$, G1 does not play an essential role in the temperature regime where spontaneous magnetization is expected to develop. The bottom figure shows \textbf{LP${}_{\mathbb{Z}_2}(D_{1331})$} bounds on $\overline{\langle s_is_{i+e_1}\rangle}$ at $p=0.2$ with and without G1 imposed over the $D_{131}$ region. In this case, we observe that G1 improves the bootstrap bounds, particularly the lower bounds. Nonetheless, the improvements are less than $\approx 10\%$ of the expected values of $\overline{\langle s_is_{i+e_1}\rangle}$.

\hspace{8pt}

\section{Envelope theorem}\label{sec:envelope}
When an optimization problem depends on external parameters, the envelope theorem describes how to obtain the derivative of the optimal value with respect to those parameters. We briefly present the ideas behind the theorem for the LP problem of interest here, which can be regarded as a special case of the more general setting discussed, e.g., in section 4.2 of \cite{Reehorst:2021ykw}.

The LP problems of interest take the following form:
\begin{eqnarray}
    &&\text{maximize } \mathbf{a}^T\mathbf{x}+b \text{ over }\mathbf{x}\in\mathbb{R}^n\nonumber
    \\
    &&\text{subject to } A(\lambda)\mathbf{x}+\mathbf{c}(\lambda)=0,~B\mathbf{x}+\mathbf{d}\geq0,
\end{eqnarray}
where $\mathbf{a}\in\mathbb{R}^n,b\in\mathbb{R},\mathbf{c}\in\mathbb{R}^m,\mathbf{d}\in\mathbb{R}^l,A\in\mathbb{R}^{m\times n}$, and $B\in\mathbb{R}^{l\times n}$. The problem depends on the external parameter $\lambda\in\mathbb{R}$ through $A(\lambda)$ and $\mathbf{c}(\lambda)$, as in the DCP, where $A(\lambda)\mathbf{x}+\mathbf{c}(\lambda)=0$ represents the invariance equations that depend on the infection rate $\lambda$.

Introducing the Lagrange multipliers $\mathbf{y}\in\mathbb{R}^m$ and $\mathbf{z}\in(\mathbb{R}_{\geq0})^l$ as dual variables, the Lagrangian is given by
\begin{equation}
    \mathcal{L}(\mathbf{x},\mathbf{y},\mathbf{z};\lambda)=\mathbf{a}^T\mathbf{x}+b+\mathbf{y}^T\left(A(\lambda)\mathbf{x}+\mathbf{c}(\lambda)\right)+\mathbf{z}^T(B\mathbf{x}+\mathbf{d}).
\end{equation}
At the primal-dual optimal point $(\mathbf{x}_*,\mathbf{y}_*,\mathbf{z}_*)$, the first-order condition and complementary slackness condition
\begin{equation}
    \mathbf{a}^T+\mathbf{y}_*^TA+\mathbf{z}_*^TB=0,~~(\mathbf{z}_*)_i(B\mathbf{x}_*+\mathbf{d})_i=0,~\forall i
\end{equation}
are satisfied. Denoting the maximum at the optimal point by $W(\lambda)=\mathbf{a}^T\mathbf{x}_*(\lambda)+b$, its derivative with respect to $\lambda$ is then given by
\begin{equation}
    {dW(\lambda)\over d\lambda}=\mathbf{a}^T\mathbf{x}_*'(\lambda)=-\mathbf{y}_*^TA\mathbf{x}_*'(\lambda)-\mathbf{z}_*^TB\mathbf{x}_*'(\lambda).
\end{equation}
The second term $\mathbf{z}_*^TB\mathbf{x}_*'(\lambda)$ vanishes due to complementary slackness. If $\left(B\mathbf{x}_*+\mathbf{d}\right)_i>0$, then $(\mathbf{z}_*)_i=0$, while if $\left(B\mathbf{x}_*+\mathbf{d}\right)_i=0$, differentiating it with respect to $\lambda$ gives $B(\mathbf{x}_*)_i'(\lambda)=0$ since $B$ and $\mathbf{d}$ are independent of $\lambda$. Differentiating the equality constraints $A(\lambda)\mathbf{x}_*(\lambda)+\mathbf{c}(\lambda)=0$ with respect to $\lambda$ gives $-A\mathbf{x}_*'(\lambda)=A'(\lambda)\mathbf{x}_*(\lambda)+\mathbf{c}'(\lambda)$, and we thus arrive at the desired envelope theorem
\begin{equation}
    {dW(\lambda)\over d\lambda}=\mathbf{y}_*^T(A'(\lambda)\mathbf{x}_*(\lambda)+\mathbf{c}'(\lambda)).
\end{equation}
Since primal-dual LP solvers provide the values of $(\mathbf{x}_*,\mathbf{y}_*,\mathbf{z}_*)$ at optimality, ${dW(\lambda)\over d\lambda}$ can be obtained straightforwardly.

\bibliography{apssamp}

\begin{thebibliography}{56}%
\makeatletter
\providecommand \@ifxundefined [1]{%
 \@ifx{#1\undefined}
}%
\providecommand \@ifnum [1]{%
 \ifnum #1\expandafter \@firstoftwo
 \else \expandafter \@secondoftwo
 \fi
}%
\providecommand \@ifx [1]{%
 \ifx #1\expandafter \@firstoftwo
 \else \expandafter \@secondoftwo
 \fi
}%
\providecommand \natexlab [1]{#1}%
\providecommand \enquote  [1]{``#1''}%
\providecommand \bibnamefont  [1]{#1}%
\providecommand \bibfnamefont [1]{#1}%
\providecommand \citenamefont [1]{#1}%
\providecommand \href@noop [0]{\@secondoftwo}%
\providecommand \href [0]{\begingroup \@sanitize@url \@href}%
\providecommand \@href[1]{\@@startlink{#1}\@@href}%
\providecommand \@@href[1]{\endgroup#1\@@endlink}%
\providecommand \@sanitize@url [0]{\catcode `\\12\catcode `\$12\catcode `\&12\catcode `\#12\catcode `\^12\catcode `\_12\catcode `\%12\relax}%
\providecommand \@@startlink[1]{}%
\providecommand \@@endlink[0]{}%
\providecommand \url  [0]{\begingroup\@sanitize@url \@url }%
\providecommand \@url [1]{\endgroup\@href {#1}{\urlprefix }}%
\providecommand \urlprefix  [0]{URL }%
\providecommand \Eprint [0]{\href }%
\providecommand \doibase [0]{https://doi.org/}%
\providecommand \selectlanguage [0]{\@gobble}%
\providecommand \bibinfo  [0]{\@secondoftwo}%
\providecommand \bibfield  [0]{\@secondoftwo}%
\providecommand \translation [1]{[#1]}%
\providecommand \BibitemOpen [0]{}%
\providecommand \bibitemStop [0]{}%
\providecommand \bibitemNoStop [0]{.\EOS\space}%
\providecommand \EOS [0]{\spacefactor3000\relax}%
\providecommand \BibitemShut  [1]{\csname bibitem#1\endcsname}%
\let\auto@bib@innerbib\@empty
\bibitem [{\citenamefont {{Anderson}}(1958)}]{1958PhRv..109.1492A}%
  \BibitemOpen
  \bibfield  {author} {\bibinfo {author} {\bibfnamefont {P.~W.}\ \bibnamefont {{Anderson}}},\ }\bibfield  {title} {\bibinfo {title} {{Absence of Diffusion in Certain Random Lattices}},\ }\href {https://doi.org/10.1103/PhysRev.109.1492} {\bibfield  {journal} {\bibinfo  {journal} {Physical Review}\ }\textbf {\bibinfo {volume} {109}},\ \bibinfo {pages} {1492} (\bibinfo {year} {1958})}\BibitemShut {NoStop}%
\bibitem [{\citenamefont {Harris}(1974{\natexlab{a}})}]{ABHarris_1974}%
  \BibitemOpen
  \bibfield  {author} {\bibinfo {author} {\bibfnamefont {A.~B.}\ \bibnamefont {Harris}},\ }\bibfield  {title} {\bibinfo {title} {Effect of random defects on the critical behaviour of ising models},\ }\href {https://doi.org/10.1088/0022-3719/7/9/009} {\bibfield  {journal} {\bibinfo  {journal} {Journal of Physics C: Solid State Physics}\ }\textbf {\bibinfo {volume} {7}},\ \bibinfo {pages} {1671} (\bibinfo {year} {1974}{\natexlab{a}})}\BibitemShut {NoStop}%
\bibitem [{\citenamefont {Imry}\ and\ \citenamefont {Ma}(1975)}]{PhysRevLett.35.1399}%
  \BibitemOpen
  \bibfield  {author} {\bibinfo {author} {\bibfnamefont {Y.}~\bibnamefont {Imry}}\ and\ \bibinfo {author} {\bibfnamefont {S.-k.}\ \bibnamefont {Ma}},\ }\bibfield  {title} {\bibinfo {title} {Random-field instability of the ordered state of continuous symmetry},\ }\href {https://doi.org/10.1103/PhysRevLett.35.1399} {\bibfield  {journal} {\bibinfo  {journal} {Phys. Rev. Lett.}\ }\textbf {\bibinfo {volume} {35}},\ \bibinfo {pages} {1399} (\bibinfo {year} {1975})}\BibitemShut {NoStop}%
\bibitem [{\citenamefont {Aizenman}\ and\ \citenamefont {Wehr}(1989)}]{PhysRevLett.62.2503}%
  \BibitemOpen
  \bibfield  {author} {\bibinfo {author} {\bibfnamefont {M.}~\bibnamefont {Aizenman}}\ and\ \bibinfo {author} {\bibfnamefont {J.}~\bibnamefont {Wehr}},\ }\bibfield  {title} {\bibinfo {title} {Rounding of first-order phase transitions in systems with quenched disorder},\ }\href {https://doi.org/10.1103/PhysRevLett.62.2503} {\bibfield  {journal} {\bibinfo  {journal} {Phys. Rev. Lett.}\ }\textbf {\bibinfo {volume} {62}},\ \bibinfo {pages} {2503} (\bibinfo {year} {1989})}\BibitemShut {NoStop}%
\bibitem [{\citenamefont {Aizenman}\ and\ \citenamefont {Wehr}(1990)}]{Aizenman1990}%
  \BibitemOpen
  \bibfield  {author} {\bibinfo {author} {\bibfnamefont {M.}~\bibnamefont {Aizenman}}\ and\ \bibinfo {author} {\bibfnamefont {J.}~\bibnamefont {Wehr}},\ }\bibfield  {title} {\bibinfo {title} {Rounding effects of quenched randomness on first-order phase transitions},\ }\href {https://doi.org/10.1007/BF02096933} {\bibfield  {journal} {\bibinfo  {journal} {Communications in Mathematical Physics}\ }\textbf {\bibinfo {volume} {130}},\ \bibinfo {pages} {489} (\bibinfo {year} {1990})}\BibitemShut {NoStop}%
\bibitem [{\citenamefont {Edwards}\ and\ \citenamefont {Anderson}(1975)}]{SFEdwards_1975}%
  \BibitemOpen
  \bibfield  {author} {\bibinfo {author} {\bibfnamefont {S.~F.}\ \bibnamefont {Edwards}}\ and\ \bibinfo {author} {\bibfnamefont {P.~W.}\ \bibnamefont {Anderson}},\ }\bibfield  {title} {\bibinfo {title} {Theory of spin glasses},\ }\href {https://doi.org/10.1088/0305-4608/5/5/017} {\bibfield  {journal} {\bibinfo  {journal} {Journal of Physics F: Metal Physics}\ }\textbf {\bibinfo {volume} {5}},\ \bibinfo {pages} {965} (\bibinfo {year} {1975})}\BibitemShut {NoStop}%
\bibitem [{\citenamefont {Sherrington}\ and\ \citenamefont {Kirkpatrick}(1975)}]{PhysRevLett.35.1792}%
  \BibitemOpen
  \bibfield  {author} {\bibinfo {author} {\bibfnamefont {D.}~\bibnamefont {Sherrington}}\ and\ \bibinfo {author} {\bibfnamefont {S.}~\bibnamefont {Kirkpatrick}},\ }\bibfield  {title} {\bibinfo {title} {Solvable model of a spin-glass},\ }\href {https://doi.org/10.1103/PhysRevLett.35.1792} {\bibfield  {journal} {\bibinfo  {journal} {Phys. Rev. Lett.}\ }\textbf {\bibinfo {volume} {35}},\ \bibinfo {pages} {1792} (\bibinfo {year} {1975})}\BibitemShut {NoStop}%
\bibitem [{\citenamefont {M{\'e}zard}\ \emph {et~al.}(1987)\citenamefont {M{\'e}zard}, \citenamefont {Parisi},\ and\ \citenamefont {Virasoro}}]{Mezard1987}%
  \BibitemOpen
  \bibfield  {author} {\bibinfo {author} {\bibfnamefont {M.}~\bibnamefont {M{\'e}zard}}, \bibinfo {author} {\bibfnamefont {G.}~\bibnamefont {Parisi}},\ and\ \bibinfo {author} {\bibfnamefont {M.~A.}\ \bibnamefont {Virasoro}},\ }\href {https://doi.org/10.1142/0271} {\emph {\bibinfo {title} {Spin Glass Theory and Beyond: An Introduction to the Replica Method and Its Applications}}},\ \bibinfo {series} {World Scientific Lecture Notes in Physics}, Vol.~\bibinfo {volume} {9}\ (\bibinfo  {publisher} {World Scientific Publishing},\ \bibinfo {address} {Singapore},\ \bibinfo {year} {1987})\BibitemShut {NoStop}%
\bibitem [{\citenamefont {Jain}(1992)}]{Jain1992}%
  \BibitemOpen
  \bibfield  {author} {\bibinfo {author} {\bibfnamefont {S.}~\bibnamefont {Jain}},\ }\href {https://doi.org/10.1142/0654} {\emph {\bibinfo {title} {Monte Carlo Simulations of Disordered Systems}}}\ (\bibinfo  {publisher} {World Scientific Publishing},\ \bibinfo {address} {Singapore},\ \bibinfo {year} {1992})\BibitemShut {NoStop}%
\bibitem [{\citenamefont {Griffiths}(1969)}]{PhysRevLett.23.17}%
  \BibitemOpen
  \bibfield  {author} {\bibinfo {author} {\bibfnamefont {R.~B.}\ \bibnamefont {Griffiths}},\ }\bibfield  {title} {\bibinfo {title} {Nonanalytic behavior above the critical point in a random ising ferromagnet},\ }\href {https://doi.org/10.1103/PhysRevLett.23.17} {\bibfield  {journal} {\bibinfo  {journal} {Phys. Rev. Lett.}\ }\textbf {\bibinfo {volume} {23}},\ \bibinfo {pages} {17} (\bibinfo {year} {1969})}\BibitemShut {NoStop}%
\bibitem [{\citenamefont {Cho}\ \emph {et~al.}(2022)\citenamefont {Cho}, \citenamefont {Gabai}, \citenamefont {Lin}, \citenamefont {Rodriguez}, \citenamefont {Sandor},\ and\ \citenamefont {Yin}}]{Cho:2022lcj}%
  \BibitemOpen
  \bibfield  {author} {\bibinfo {author} {\bibfnamefont {M.}~\bibnamefont {Cho}}, \bibinfo {author} {\bibfnamefont {B.}~\bibnamefont {Gabai}}, \bibinfo {author} {\bibfnamefont {Y.-H.}\ \bibnamefont {Lin}}, \bibinfo {author} {\bibfnamefont {V.~A.}\ \bibnamefont {Rodriguez}}, \bibinfo {author} {\bibfnamefont {J.}~\bibnamefont {Sandor}},\ and\ \bibinfo {author} {\bibfnamefont {X.}~\bibnamefont {Yin}},\ }\bibfield  {title} {\bibinfo {title} {{Bootstrapping the Ising Model on the Lattice}},\ }\href@noop {} {\  (\bibinfo {year} {2022})},\ \Eprint {https://arxiv.org/abs/2206.12538} {arXiv:2206.12538 [hep-th]} \BibitemShut {NoStop}%
\bibitem [{\citenamefont {Cho}\ and\ \citenamefont {Sun}(2023)}]{Cho:2023ulr}%
  \BibitemOpen
  \bibfield  {author} {\bibinfo {author} {\bibfnamefont {M.}~\bibnamefont {Cho}}\ and\ \bibinfo {author} {\bibfnamefont {X.}~\bibnamefont {Sun}},\ }\bibfield  {title} {\bibinfo {title} {{Bootstrap, Markov Chain Monte Carlo, and LP/SDP hierarchy for the lattice Ising model}},\ }\href {https://doi.org/10.1007/JHEP11(2023)047} {\bibfield  {journal} {\bibinfo  {journal} {JHEP}\ }\textbf {\bibinfo {volume} {11}},\ \bibinfo {pages} {047}},\ \Eprint {https://arxiv.org/abs/2309.01016} {arXiv:2309.01016 [hep-th]} \BibitemShut {NoStop}%
\bibitem [{\citenamefont {Cho}(2025)}]{Cho:2025dgc}%
  \BibitemOpen
  \bibfield  {author} {\bibinfo {author} {\bibfnamefont {M.}~\bibnamefont {Cho}},\ }\bibfield  {title} {\bibinfo {title} {{Bootstrapping nonequilibrium stochastic processes}},\ }\href {https://doi.org/10.21468/SciPostPhys.19.5.124} {\bibfield  {journal} {\bibinfo  {journal} {SciPost Phys.}\ }\textbf {\bibinfo {volume} {19}},\ \bibinfo {pages} {124} (\bibinfo {year} {2025})},\ \Eprint {https://arxiv.org/abs/2505.13609} {arXiv:2505.13609 [cond-mat.stat-mech]} \BibitemShut {NoStop}%
\bibitem [{\citenamefont {Önder}\ \emph {et~al.}(pear)\citenamefont {Önder}, \citenamefont {Scheer}, \citenamefont {Cho},\ and\ \citenamefont {Khalaf}}]{unpubl1}%
  \BibitemOpen
  \bibfield  {author} {\bibinfo {author} {\bibfnamefont {Y.}~\bibnamefont {Önder}}, \bibinfo {author} {\bibfnamefont {M.~G.}\ \bibnamefont {Scheer}}, \bibinfo {author} {\bibfnamefont {M.}~\bibnamefont {Cho}},\ and\ \bibinfo {author} {\bibfnamefont {E.}~\bibnamefont {Khalaf}},\ }\bibfield  {title} {\bibinfo {title} {Bootstrapping disordered quantum systems}} (\bibinfo {year} {to appear})\BibitemShut {NoStop}%
\bibitem [{\citenamefont {Bieche}\ \emph {et~al.}(1980)\citenamefont {Bieche}, \citenamefont {Uhry}, \citenamefont {Maynard},\ and\ \citenamefont {Rammal}}]{LBieche_1980}%
  \BibitemOpen
  \bibfield  {author} {\bibinfo {author} {\bibfnamefont {L.}~\bibnamefont {Bieche}}, \bibinfo {author} {\bibfnamefont {J.~P.}\ \bibnamefont {Uhry}}, \bibinfo {author} {\bibfnamefont {R.}~\bibnamefont {Maynard}},\ and\ \bibinfo {author} {\bibfnamefont {R.}~\bibnamefont {Rammal}},\ }\bibfield  {title} {\bibinfo {title} {On the ground states of the frustration model of a spin glass by a matching method of graph theory},\ }\href {https://doi.org/10.1088/0305-4470/13/8/005} {\bibfield  {journal} {\bibinfo  {journal} {Journal of Physics A: Mathematical and General}\ }\textbf {\bibinfo {volume} {13}},\ \bibinfo {pages} {2553} (\bibinfo {year} {1980})}\BibitemShut {NoStop}%
\bibitem [{\citenamefont {Barahona}(1982)}]{FBarahona_1982}%
  \BibitemOpen
  \bibfield  {author} {\bibinfo {author} {\bibfnamefont {F.}~\bibnamefont {Barahona}},\ }\bibfield  {title} {\bibinfo {title} {On the computational complexity of ising spin glass models},\ }\href {https://doi.org/10.1088/0305-4470/15/10/028} {\bibfield  {journal} {\bibinfo  {journal} {Journal of Physics A: Mathematical and General}\ }\textbf {\bibinfo {volume} {15}},\ \bibinfo {pages} {3241} (\bibinfo {year} {1982})}\BibitemShut {NoStop}%
\bibitem [{\citenamefont {Barahona}\ \emph {et~al.}(1989)\citenamefont {Barahona}, \citenamefont {J{\"u}nger},\ and\ \citenamefont {Reinelt}}]{Barahona1989}%
  \BibitemOpen
  \bibfield  {author} {\bibinfo {author} {\bibfnamefont {F.}~\bibnamefont {Barahona}}, \bibinfo {author} {\bibfnamefont {M.}~\bibnamefont {J{\"u}nger}},\ and\ \bibinfo {author} {\bibfnamefont {G.}~\bibnamefont {Reinelt}},\ }\bibfield  {title} {\bibinfo {title} {Experiments in quadratic 0–1 programming},\ }\href {https://doi.org/10.1007/BF01587084} {\bibfield  {journal} {\bibinfo  {journal} {Mathematical Programming}\ }\textbf {\bibinfo {volume} {44}},\ \bibinfo {pages} {127} (\bibinfo {year} {1989})}\BibitemShut {NoStop}%
\bibitem [{\citenamefont {Goemans}\ and\ \citenamefont {Williamson}(1995)}]{Goemans1995ImprovedAA}%
  \BibitemOpen
  \bibfield  {author} {\bibinfo {author} {\bibfnamefont {M.~X.}\ \bibnamefont {Goemans}}\ and\ \bibinfo {author} {\bibfnamefont {D.~P.}\ \bibnamefont {Williamson}},\ }\bibfield  {title} {\bibinfo {title} {Improved approximation algorithms for maximum cut and satisfiability problems using semidefinite programming},\ }\href@noop {} {\bibfield  {journal} {\bibinfo  {journal} {J. ACM}\ }\textbf {\bibinfo {volume} {42}},\ \bibinfo {pages} {1115} (\bibinfo {year} {1995})}\BibitemShut {NoStop}%
\bibitem [{Note1()}]{Note1}%
  \BibitemOpen
  \bibinfo {note} {This excludes systems described by signed or complex measures that appear in certain lattice gauge theories, for which positivity (\ref {eq:positivity}) does not apply.}\BibitemShut {Stop}%
\bibitem [{Note2()}]{Note2}%
  \BibitemOpen
  \bibinfo {note} {We also remark that a closely related idea of considering the joint space of random couplings and quantum states has been applied to the tensor network representation of quantum many-body states for quantum systems with quenched disorder in \cite {PhysRevLett.95.140501,SciPostPhys.6.3.031,Vervoort:2025auj}.}\BibitemShut {Stop}%
\bibitem [{\citenamefont {Dobruschin}(1968)}]{doi:10.1137/1113026}%
  \BibitemOpen
  \bibfield  {author} {\bibinfo {author} {\bibfnamefont {P.~L.}\ \bibnamefont {Dobruschin}},\ }\bibfield  {title} {\bibinfo {title} {The description of a random field by means of conditional probabilities and conditions of its regularity},\ }\href {https://doi.org/10.1137/1113026} {\bibfield  {journal} {\bibinfo  {journal} {Theory of Probability \& Its Applications}\ }\textbf {\bibinfo {volume} {13}},\ \bibinfo {pages} {197} (\bibinfo {year} {1968})},\ \Eprint {https://arxiv.org/abs/https://doi.org/10.1137/1113026} {https://doi.org/10.1137/1113026} \BibitemShut {NoStop}%
\bibitem [{\citenamefont {Dobrushin}(1968)}]{Dobrushin1968}%
  \BibitemOpen
  \bibfield  {author} {\bibinfo {author} {\bibfnamefont {R.~L.}\ \bibnamefont {Dobrushin}},\ }\bibfield  {title} {\bibinfo {title} {The problem of uniqueness of a gibbsian random field and the problem of phase transitions},\ }\href {https://doi.org/10.1007/BF01075682} {\bibfield  {journal} {\bibinfo  {journal} {Functional Analysis and Its Applications}\ }\textbf {\bibinfo {volume} {2}},\ \bibinfo {pages} {302} (\bibinfo {year} {1968})}\BibitemShut {NoStop}%
\bibitem [{\citenamefont {Lanford}\ and\ \citenamefont {Ruelle}(1969)}]{Lanford1969}%
  \BibitemOpen
  \bibfield  {author} {\bibinfo {author} {\bibfnamefont {O.~E.}\ \bibnamefont {Lanford}}\ and\ \bibinfo {author} {\bibfnamefont {D.}~\bibnamefont {Ruelle}},\ }\bibfield  {title} {\bibinfo {title} {Observables at infinity and states with short range correlations in statistical mechanics},\ }\href {https://doi.org/10.1007/BF01645487} {\bibfield  {journal} {\bibinfo  {journal} {Communications in Mathematical Physics}\ }\textbf {\bibinfo {volume} {13}},\ \bibinfo {pages} {194} (\bibinfo {year} {1969})}\BibitemShut {NoStop}%
\bibitem [{\citenamefont {{MOSEK ApS}}(2026)}]{mosek2026}%
  \BibitemOpen
  \bibfield  {author} {\bibinfo {author} {\bibnamefont {{MOSEK ApS}}},\ }\href {https://docs.mosek.com/latest/toolbox/index.html} {\emph {\bibinfo {title} {The MOSEK Optimization Toolbox for MATLAB Manual}}} (\bibinfo {year} {2026}),\ \bibinfo {note} {version 11.2}\BibitemShut {NoStop}%
\bibitem [{\citenamefont {Brout}(1959)}]{PhysRev.115.824}%
  \BibitemOpen
  \bibfield  {author} {\bibinfo {author} {\bibfnamefont {R.}~\bibnamefont {Brout}},\ }\bibfield  {title} {\bibinfo {title} {Statistical mechanical theory of a random ferromagnetic system},\ }\href {https://doi.org/10.1103/PhysRev.115.824} {\bibfield  {journal} {\bibinfo  {journal} {Phys. Rev.}\ }\textbf {\bibinfo {volume} {115}},\ \bibinfo {pages} {824} (\bibinfo {year} {1959})}\BibitemShut {NoStop}%
\bibitem [{\citenamefont {Elliott}\ \emph {et~al.}(1960)\citenamefont {Elliott}, \citenamefont {Heap}, \citenamefont {Morgan},\ and\ \citenamefont {Rushbrooke}}]{PhysRevLett.5.366}%
  \BibitemOpen
  \bibfield  {author} {\bibinfo {author} {\bibfnamefont {R.~J.}\ \bibnamefont {Elliott}}, \bibinfo {author} {\bibfnamefont {B.~R.}\ \bibnamefont {Heap}}, \bibinfo {author} {\bibfnamefont {D.~J.}\ \bibnamefont {Morgan}},\ and\ \bibinfo {author} {\bibfnamefont {G.~S.}\ \bibnamefont {Rushbrooke}},\ }\bibfield  {title} {\bibinfo {title} {Equivalence of the critical concentrations in the ising and heisenberg models of ferromagnetism},\ }\href {https://doi.org/10.1103/PhysRevLett.5.366} {\bibfield  {journal} {\bibinfo  {journal} {Phys. Rev. Lett.}\ }\textbf {\bibinfo {volume} {5}},\ \bibinfo {pages} {366} (\bibinfo {year} {1960})}\BibitemShut {NoStop}%
\bibitem [{\citenamefont {Bramson}\ \emph {et~al.}(1991)\citenamefont {Bramson}, \citenamefont {Durrett},\ and\ \citenamefont {Schonmann}}]{10.1214/aop/1176990331}%
  \BibitemOpen
  \bibfield  {author} {\bibinfo {author} {\bibfnamefont {M.}~\bibnamefont {Bramson}}, \bibinfo {author} {\bibfnamefont {R.}~\bibnamefont {Durrett}},\ and\ \bibinfo {author} {\bibfnamefont {R.~H.}\ \bibnamefont {Schonmann}},\ }\bibfield  {title} {\bibinfo {title} {{The Contact Processes in a Random Environment}},\ }\href {https://doi.org/10.1214/aop/1176990331} {\bibfield  {journal} {\bibinfo  {journal} {The Annals of Probability}\ }\textbf {\bibinfo {volume} {19}},\ \bibinfo {pages} {960 } (\bibinfo {year} {1991})}\BibitemShut {NoStop}%
\bibitem [{\citenamefont {Harris}(1974{\natexlab{b}})}]{10.1214/aop/1176996493}%
  \BibitemOpen
  \bibfield  {author} {\bibinfo {author} {\bibfnamefont {T.~E.}\ \bibnamefont {Harris}},\ }\bibfield  {title} {\bibinfo {title} {{Contact Interactions on a Lattice}},\ }\href {https://doi.org/10.1214/aop/1176996493} {\bibfield  {journal} {\bibinfo  {journal} {The Annals of Probability}\ }\textbf {\bibinfo {volume} {2}},\ \bibinfo {pages} {969 } (\bibinfo {year} {1974}{\natexlab{b}})}\BibitemShut {NoStop}%
\bibitem [{\citenamefont {Liggett}(1985)}]{liggett1985interacting}%
  \BibitemOpen
  \bibfield  {author} {\bibinfo {author} {\bibfnamefont {T.~M.}\ \bibnamefont {Liggett}},\ }\href {https://doi.org/10.1007/978-1-4613-8542-4} {\emph {\bibinfo {title} {Interacting Particle Systems}}},\ \bibinfo {series} {Grundlehren der mathematischen Wissenschaften}, Vol.\ \bibinfo {volume} {276}\ (\bibinfo  {publisher} {Springer-Verlag},\ \bibinfo {address} {New York},\ \bibinfo {year} {1985})\BibitemShut {NoStop}%
\bibitem [{\citenamefont {Janssen}(1981)}]{Janssen1981}%
  \BibitemOpen
  \bibfield  {author} {\bibinfo {author} {\bibfnamefont {H.~K.}\ \bibnamefont {Janssen}},\ }\bibfield  {title} {\bibinfo {title} {On the nonequilibrium phase transition in reaction-diffusion systems with an absorbing stationary state},\ }\href {https://doi.org/10.1007/BF01319549} {\bibfield  {journal} {\bibinfo  {journal} {Zeitschrift f{\"u}r Physik B Condensed Matter}\ }\textbf {\bibinfo {volume} {42}},\ \bibinfo {pages} {151} (\bibinfo {year} {1981})}\BibitemShut {NoStop}%
\bibitem [{\citenamefont {Hooyberghs}\ \emph {et~al.}(2003)\citenamefont {Hooyberghs}, \citenamefont {Igl\'oi},\ and\ \citenamefont {Vanderzande}}]{PhysRevLett.90.100601}%
  \BibitemOpen
  \bibfield  {author} {\bibinfo {author} {\bibfnamefont {J.}~\bibnamefont {Hooyberghs}}, \bibinfo {author} {\bibfnamefont {F.}~\bibnamefont {Igl\'oi}},\ and\ \bibinfo {author} {\bibfnamefont {C.}~\bibnamefont {Vanderzande}},\ }\bibfield  {title} {\bibinfo {title} {Strong disorder fixed point in absorbing-state phase transitions},\ }\href {https://doi.org/10.1103/PhysRevLett.90.100601} {\bibfield  {journal} {\bibinfo  {journal} {Phys. Rev. Lett.}\ }\textbf {\bibinfo {volume} {90}},\ \bibinfo {pages} {100601} (\bibinfo {year} {2003})}\BibitemShut {NoStop}%
\bibitem [{\citenamefont {Kramers}\ and\ \citenamefont {Wannier}(1941{\natexlab{a}})}]{PhysRev.60.252}%
  \BibitemOpen
  \bibfield  {author} {\bibinfo {author} {\bibfnamefont {H.~A.}\ \bibnamefont {Kramers}}\ and\ \bibinfo {author} {\bibfnamefont {G.~H.}\ \bibnamefont {Wannier}},\ }\bibfield  {title} {\bibinfo {title} {Statistics of the two-dimensional ferromagnet. part i},\ }\href {https://doi.org/10.1103/PhysRev.60.252} {\bibfield  {journal} {\bibinfo  {journal} {Phys. Rev.}\ }\textbf {\bibinfo {volume} {60}},\ \bibinfo {pages} {252} (\bibinfo {year} {1941}{\natexlab{a}})}\BibitemShut {NoStop}%
\bibitem [{\citenamefont {Kramers}\ and\ \citenamefont {Wannier}(1941{\natexlab{b}})}]{PhysRev.60.263}%
  \BibitemOpen
  \bibfield  {author} {\bibinfo {author} {\bibfnamefont {H.~A.}\ \bibnamefont {Kramers}}\ and\ \bibinfo {author} {\bibfnamefont {G.~H.}\ \bibnamefont {Wannier}},\ }\bibfield  {title} {\bibinfo {title} {Statistics of the two-dimensional ferromagnet. part ii},\ }\href {https://doi.org/10.1103/PhysRev.60.263} {\bibfield  {journal} {\bibinfo  {journal} {Phys. Rev.}\ }\textbf {\bibinfo {volume} {60}},\ \bibinfo {pages} {263} (\bibinfo {year} {1941}{\natexlab{b}})}\BibitemShut {NoStop}%
\bibitem [{\citenamefont {{Ben Al{\`\i} Zinati}}\ \emph {et~al.}(2026)\citenamefont {{Ben Al{\`\i} Zinati}}, \citenamefont {{Gori}},\ and\ \citenamefont {{Codello}}}]{2026arXiv260321303B}%
  \BibitemOpen
  \bibfield  {author} {\bibinfo {author} {\bibfnamefont {R.}~\bibnamefont {{Ben Al{\`\i} Zinati}}}, \bibinfo {author} {\bibfnamefont {G.}~\bibnamefont {{Gori}}},\ and\ \bibinfo {author} {\bibfnamefont {A.}~\bibnamefont {{Codello}}},\ }\bibfield  {title} {\bibinfo {title} {{The phase boundary of the random site Ising model}},\ }\href {https://doi.org/10.48550/arXiv.2603.21303} {\bibfield  {journal} {\bibinfo  {journal} {arXiv e-prints}\ ,\ \bibinfo {eid} {arXiv:2603.21303}} (\bibinfo {year} {2026})},\ \Eprint {https://arxiv.org/abs/2603.21303} {arXiv:2603.21303 [cond-mat.stat-mech]} \BibitemShut {NoStop}%
\bibitem [{\citenamefont {Griffiths}(1967{\natexlab{a}})}]{10.1063/1.1705219}%
  \BibitemOpen
  \bibfield  {author} {\bibinfo {author} {\bibfnamefont {R.~B.}\ \bibnamefont {Griffiths}},\ }\bibfield  {title} {\bibinfo {title} {Correlations in ising ferromagnets. i},\ }\href {https://doi.org/10.1063/1.1705219} {\bibfield  {journal} {\bibinfo  {journal} {Journal of Mathematical Physics}\ }\textbf {\bibinfo {volume} {8}},\ \bibinfo {pages} {478} (\bibinfo {year} {1967}{\natexlab{a}})}\BibitemShut {NoStop}%
\bibitem [{\citenamefont {Griffiths}(1967{\natexlab{b}})}]{10.1063/1.1705220}%
  \BibitemOpen
  \bibfield  {author} {\bibinfo {author} {\bibfnamefont {R.~B.}\ \bibnamefont {Griffiths}},\ }\bibfield  {title} {\bibinfo {title} {Correlations in ising ferromagnets. ii. external magnetic fields},\ }\href {https://doi.org/10.1063/1.1705220} {\bibfield  {journal} {\bibinfo  {journal} {Journal of Mathematical Physics}\ }\textbf {\bibinfo {volume} {8}},\ \bibinfo {pages} {484} (\bibinfo {year} {1967}{\natexlab{b}})}\BibitemShut {NoStop}%
\bibitem [{\citenamefont {Griffiths}(1967{\natexlab{c}})}]{Griffiths1967}%
  \BibitemOpen
  \bibfield  {author} {\bibinfo {author} {\bibfnamefont {R.~B.}\ \bibnamefont {Griffiths}},\ }\bibfield  {title} {\bibinfo {title} {Correlations in ising ferromagnets. {III}},\ }\href {https://doi.org/10.1007/BF01654128} {\bibfield  {journal} {\bibinfo  {journal} {Communications in Mathematical Physics}\ }\textbf {\bibinfo {volume} {6}},\ \bibinfo {pages} {121} (\bibinfo {year} {1967}{\natexlab{c}})}\BibitemShut {NoStop}%
\bibitem [{\citenamefont {{Marro}}\ and\ \citenamefont {{Dickman}}(2005)}]{2005nptl.book.....M}%
  \BibitemOpen
  \bibfield  {author} {\bibinfo {author} {\bibfnamefont {J.}~\bibnamefont {{Marro}}}\ and\ \bibinfo {author} {\bibfnamefont {R.}~\bibnamefont {{Dickman}}},\ }\href@noop {} {\emph {\bibinfo {title} {{Nonequilibrium Phase Transitions in Lattice Models}}}}\ (\bibinfo {year} {2005})\BibitemShut {NoStop}%
\bibitem [{\citenamefont {{Vojta}}\ and\ \citenamefont {{Dickison}}(2005)}]{2005PhRvE..72c6126V}%
  \BibitemOpen
  \bibfield  {author} {\bibinfo {author} {\bibfnamefont {T.}~\bibnamefont {{Vojta}}}\ and\ \bibinfo {author} {\bibfnamefont {M.}~\bibnamefont {{Dickison}}},\ }\bibfield  {title} {\bibinfo {title} {{Critical behavior and Griffiths effects in the disordered contact process}},\ }\href {https://doi.org/10.1103/PhysRevE.72.036126} {\bibfield  {journal} {\bibinfo  {journal} {\pre}\ }\textbf {\bibinfo {volume} {72}},\ \bibinfo {eid} {036126} (\bibinfo {year} {2005})},\ \Eprint {https://arxiv.org/abs/cond-mat/0505354} {arXiv:cond-mat/0505354 [cond-mat.stat-mech]} \BibitemShut {NoStop}%
\bibitem [{\citenamefont {Inc.}()}]{Mathematica}%
  \BibitemOpen
  \bibfield  {author} {\bibinfo {author} {\bibfnamefont {W.~R.}\ \bibnamefont {Inc.}},\ }\href {https://www.wolfram.com/mathematica} {\bibinfo {title} {Mathematica, {V}ersion 14.3}},\ \bibinfo {note} {champaign, IL, 2026}\BibitemShut {NoStop}%
\bibitem [{\citenamefont {{Hinrichsen}}(2000)}]{2000AdPhy..49..815H}%
  \BibitemOpen
  \bibfield  {author} {\bibinfo {author} {\bibfnamefont {H.}~\bibnamefont {{Hinrichsen}}},\ }\bibfield  {title} {\bibinfo {title} {{Non-equilibrium critical phenomena and phase transitions into absorbing states}},\ }\href {https://doi.org/10.1080/00018730050198152} {\bibfield  {journal} {\bibinfo  {journal} {Advances in Physics}\ }\textbf {\bibinfo {volume} {49}},\ \bibinfo {pages} {815} (\bibinfo {year} {2000})},\ \Eprint {https://arxiv.org/abs/cond-mat/0001070} {arXiv:cond-mat/0001070 [cond-mat.stat-mech]} \BibitemShut {NoStop}%
\bibitem [{\citenamefont {Fisher}(1992)}]{PhysRevLett.69.534}%
  \BibitemOpen
  \bibfield  {author} {\bibinfo {author} {\bibfnamefont {D.~S.}\ \bibnamefont {Fisher}},\ }\bibfield  {title} {\bibinfo {title} {Random transverse field ising spin chains},\ }\href {https://doi.org/10.1103/PhysRevLett.69.534} {\bibfield  {journal} {\bibinfo  {journal} {Phys. Rev. Lett.}\ }\textbf {\bibinfo {volume} {69}},\ \bibinfo {pages} {534} (\bibinfo {year} {1992})}\BibitemShut {NoStop}%
\bibitem [{\citenamefont {Kull}\ \emph {et~al.}(2024)\citenamefont {Kull}, \citenamefont {Schuch}, \citenamefont {Dive},\ and\ \citenamefont {Navascu{\'e}s}}]{Kull:2022wof}%
  \BibitemOpen
  \bibfield  {author} {\bibinfo {author} {\bibfnamefont {I.}~\bibnamefont {Kull}}, \bibinfo {author} {\bibfnamefont {N.}~\bibnamefont {Schuch}}, \bibinfo {author} {\bibfnamefont {B.}~\bibnamefont {Dive}},\ and\ \bibinfo {author} {\bibfnamefont {M.}~\bibnamefont {Navascu{\'e}s}},\ }\bibfield  {title} {\bibinfo {title} {{Lower Bounds on Ground-State Energies of Local Hamiltonians through the Renormalization Group}},\ }\href {https://doi.org/10.1103/PhysRevX.14.021008} {\bibfield  {journal} {\bibinfo  {journal} {Phys. Rev. X}\ }\textbf {\bibinfo {volume} {14}},\ \bibinfo {pages} {021008} (\bibinfo {year} {2024})},\ \Eprint {https://arxiv.org/abs/2212.03014} {arXiv:2212.03014 [quant-ph]} \BibitemShut {NoStop}%
\bibitem [{\citenamefont {Cho}\ \emph {et~al.}(2026)\citenamefont {Cho}, \citenamefont {Nancarrow}, \citenamefont {Tadi{\'c}}, \citenamefont {Xin},\ and\ \citenamefont {Zheng}}]{Cho:2024owx}%
  \BibitemOpen
  \bibfield  {author} {\bibinfo {author} {\bibfnamefont {M.}~\bibnamefont {Cho}}, \bibinfo {author} {\bibfnamefont {C.~O.}\ \bibnamefont {Nancarrow}}, \bibinfo {author} {\bibfnamefont {P.}~\bibnamefont {Tadi{\'c}}}, \bibinfo {author} {\bibfnamefont {Y.}~\bibnamefont {Xin}},\ and\ \bibinfo {author} {\bibfnamefont {Z.}~\bibnamefont {Zheng}},\ }\bibfield  {title} {\bibinfo {title} {{Coarse-grained bootstrap of quantum many-body systems}},\ }\href {https://doi.org/10.1007/JHEP02(2026)222} {\bibfield  {journal} {\bibinfo  {journal} {JHEP}\ }\textbf {\bibinfo {volume} {02}},\ \bibinfo {pages} {222}},\ \Eprint {https://arxiv.org/abs/2412.07837} {arXiv:2412.07837 [hep-th]} \BibitemShut {NoStop}%
\bibitem [{\citenamefont {Mazziotti}(1998)}]{PhysRevA.57.4219}%
  \BibitemOpen
  \bibfield  {author} {\bibinfo {author} {\bibfnamefont {D.~A.}\ \bibnamefont {Mazziotti}},\ }\bibfield  {title} {\bibinfo {title} {Contracted schr\"odinger equation: Determining quantum energies and two-particle density matrices without wave functions},\ }\href {https://doi.org/10.1103/PhysRevA.57.4219} {\bibfield  {journal} {\bibinfo  {journal} {Phys. Rev. A}\ }\textbf {\bibinfo {volume} {57}},\ \bibinfo {pages} {4219} (\bibinfo {year} {1998})}\BibitemShut {NoStop}%
\bibitem [{\citenamefont {Nakata}\ \emph {et~al.}(2001)\citenamefont {Nakata}, \citenamefont {Nakatsuji}, \citenamefont {Ehara}, \citenamefont {Fukuda}, \citenamefont {Nakata},\ and\ \citenamefont {Fujisawa}}]{10.1063/1.1360199}%
  \BibitemOpen
  \bibfield  {author} {\bibinfo {author} {\bibfnamefont {M.}~\bibnamefont {Nakata}}, \bibinfo {author} {\bibfnamefont {H.}~\bibnamefont {Nakatsuji}}, \bibinfo {author} {\bibfnamefont {M.}~\bibnamefont {Ehara}}, \bibinfo {author} {\bibfnamefont {M.}~\bibnamefont {Fukuda}}, \bibinfo {author} {\bibfnamefont {K.}~\bibnamefont {Nakata}},\ and\ \bibinfo {author} {\bibfnamefont {K.}~\bibnamefont {Fujisawa}},\ }\bibfield  {title} {\bibinfo {title} {Variational calculations of fermion second-order reduced density matrices by semidefinite programming algorithm},\ }\href {https://doi.org/10.1063/1.1360199} {\bibfield  {journal} {\bibinfo  {journal} {The Journal of Chemical Physics}\ }\textbf {\bibinfo {volume} {114}},\ \bibinfo {pages} {8282} (\bibinfo {year} {2001})}\BibitemShut {NoStop}%
\bibitem [{\citenamefont {Barthel}\ and\ \citenamefont {H{\"u}bener}(2012)}]{Barthel:2012mqo}%
  \BibitemOpen
  \bibfield  {author} {\bibinfo {author} {\bibfnamefont {T.}~\bibnamefont {Barthel}}\ and\ \bibinfo {author} {\bibfnamefont {R.}~\bibnamefont {H{\"u}bener}},\ }\bibfield  {title} {\bibinfo {title} {{Solving condensed-matter ground-state problems by semidefinite relaxations}},\ }\href {https://doi.org/10.1103/PhysRevLett.108.200404} {\bibfield  {journal} {\bibinfo  {journal} {Phys. Rev. Lett.}\ }\textbf {\bibinfo {volume} {108}},\ \bibinfo {pages} {200404} (\bibinfo {year} {2012})},\ \Eprint {https://arxiv.org/abs/1106.4966} {arXiv:1106.4966 [cond-mat.str-el]} \BibitemShut {NoStop}%
\bibitem [{\citenamefont {Han}\ \emph {et~al.}(2020)\citenamefont {Han}, \citenamefont {Hartnoll},\ and\ \citenamefont {Kruthoff}}]{Han:2020bkb}%
  \BibitemOpen
  \bibfield  {author} {\bibinfo {author} {\bibfnamefont {X.}~\bibnamefont {Han}}, \bibinfo {author} {\bibfnamefont {S.~A.}\ \bibnamefont {Hartnoll}},\ and\ \bibinfo {author} {\bibfnamefont {J.}~\bibnamefont {Kruthoff}},\ }\bibfield  {title} {\bibinfo {title} {{Bootstrapping Matrix Quantum Mechanics}},\ }\href {https://doi.org/10.1103/PhysRevLett.125.041601} {\bibfield  {journal} {\bibinfo  {journal} {Phys. Rev. Lett.}\ }\textbf {\bibinfo {volume} {125}},\ \bibinfo {pages} {041601} (\bibinfo {year} {2020})},\ \Eprint {https://arxiv.org/abs/2004.10212} {arXiv:2004.10212 [hep-th]} \BibitemShut {NoStop}%
\bibitem [{\citenamefont {Nancarrow}\ and\ \citenamefont {Xin}(2023)}]{Nancarrow:2022wdr}%
  \BibitemOpen
  \bibfield  {author} {\bibinfo {author} {\bibfnamefont {C.~O.}\ \bibnamefont {Nancarrow}}\ and\ \bibinfo {author} {\bibfnamefont {Y.}~\bibnamefont {Xin}},\ }\bibfield  {title} {\bibinfo {title} {{Bootstrapping the gap in quantum spin systems}},\ }\href {https://doi.org/10.1007/JHEP08(2023)052} {\bibfield  {journal} {\bibinfo  {journal} {JHEP}\ }\textbf {\bibinfo {volume} {08}},\ \bibinfo {pages} {052}},\ \Eprint {https://arxiv.org/abs/2211.03819} {arXiv:2211.03819 [hep-th]} \BibitemShut {NoStop}%
\bibitem [{\citenamefont {Fawzi}\ \emph {et~al.}(2024)\citenamefont {Fawzi}, \citenamefont {Fawzi},\ and\ \citenamefont {Scalet}}]{Fawzi:2023fpg}%
  \BibitemOpen
  \bibfield  {author} {\bibinfo {author} {\bibfnamefont {H.}~\bibnamefont {Fawzi}}, \bibinfo {author} {\bibfnamefont {O.}~\bibnamefont {Fawzi}},\ and\ \bibinfo {author} {\bibfnamefont {S.~O.}\ \bibnamefont {Scalet}},\ }\bibfield  {title} {\bibinfo {title} {{Certified algorithms for equilibrium states of local quantum Hamiltonians}},\ }\href {https://doi.org/10.1038/s41467-024-51592-3} {\bibfield  {journal} {\bibinfo  {journal} {Nature Commun.}\ }\textbf {\bibinfo {volume} {15}},\ \bibinfo {pages} {7394} (\bibinfo {year} {2024})},\ \Eprint {https://arxiv.org/abs/2311.18706} {arXiv:2311.18706 [quant-ph]} \BibitemShut {NoStop}%
\bibitem [{\citenamefont {Gao}\ \emph {et~al.}(2025)\citenamefont {Gao}, \citenamefont {Lanzetta}, \citenamefont {Ledwith}, \citenamefont {Wang},\ and\ \citenamefont {Khalaf}}]{Gao:2024etm}%
  \BibitemOpen
  \bibfield  {author} {\bibinfo {author} {\bibfnamefont {Q.}~\bibnamefont {Gao}}, \bibinfo {author} {\bibfnamefont {R.~A.}\ \bibnamefont {Lanzetta}}, \bibinfo {author} {\bibfnamefont {P.}~\bibnamefont {Ledwith}}, \bibinfo {author} {\bibfnamefont {J.}~\bibnamefont {Wang}},\ and\ \bibinfo {author} {\bibfnamefont {E.}~\bibnamefont {Khalaf}},\ }\bibfield  {title} {\bibinfo {title} {{Bootstrapping the Quantum Hall Problem}},\ }\href {https://doi.org/10.1103/csnn-vjhn} {\bibfield  {journal} {\bibinfo  {journal} {Phys. Rev. X}\ }\textbf {\bibinfo {volume} {15}},\ \bibinfo {pages} {031034} (\bibinfo {year} {2025})},\ \Eprint {https://arxiv.org/abs/2409.10619} {arXiv:2409.10619 [cond-mat.str-el]} \BibitemShut {NoStop}%
\bibitem [{\citenamefont {Cho}\ \emph {et~al.}(2025)\citenamefont {Cho}, \citenamefont {Gabai}, \citenamefont {Sandor},\ and\ \citenamefont {Yin}}]{Cho:2024kxn}%
  \BibitemOpen
  \bibfield  {author} {\bibinfo {author} {\bibfnamefont {M.}~\bibnamefont {Cho}}, \bibinfo {author} {\bibfnamefont {B.}~\bibnamefont {Gabai}}, \bibinfo {author} {\bibfnamefont {J.}~\bibnamefont {Sandor}},\ and\ \bibinfo {author} {\bibfnamefont {X.}~\bibnamefont {Yin}},\ }\bibfield  {title} {\bibinfo {title} {{Thermal bootstrap of matrix quantum mechanics}},\ }\href {https://doi.org/10.1007/JHEP04(2025)186} {\bibfield  {journal} {\bibinfo  {journal} {JHEP}\ }\textbf {\bibinfo {volume} {04}},\ \bibinfo {pages} {186}},\ \Eprint {https://arxiv.org/abs/2410.04262} {arXiv:2410.04262 [hep-th]} \BibitemShut {NoStop}%
\bibitem [{\citenamefont {Reehorst}\ \emph {et~al.}(2021)\citenamefont {Reehorst}, \citenamefont {Rychkov}, \citenamefont {Simmons-Duffin}, \citenamefont {Sirois}, \citenamefont {Su},\ and\ \citenamefont {van Rees}}]{Reehorst:2021ykw}%
  \BibitemOpen
  \bibfield  {author} {\bibinfo {author} {\bibfnamefont {M.}~\bibnamefont {Reehorst}}, \bibinfo {author} {\bibfnamefont {S.}~\bibnamefont {Rychkov}}, \bibinfo {author} {\bibfnamefont {D.}~\bibnamefont {Simmons-Duffin}}, \bibinfo {author} {\bibfnamefont {B.}~\bibnamefont {Sirois}}, \bibinfo {author} {\bibfnamefont {N.}~\bibnamefont {Su}},\ and\ \bibinfo {author} {\bibfnamefont {B.}~\bibnamefont {van Rees}},\ }\bibfield  {title} {\bibinfo {title} {{Navigator Function for the Conformal Bootstrap}},\ }\href {https://doi.org/10.21468/SciPostPhys.11.3.072} {\bibfield  {journal} {\bibinfo  {journal} {SciPost Phys.}\ }\textbf {\bibinfo {volume} {11}},\ \bibinfo {pages} {072} (\bibinfo {year} {2021})},\ \Eprint {https://arxiv.org/abs/2104.09518} {arXiv:2104.09518 [hep-th]} \BibitemShut {NoStop}%
\bibitem [{\citenamefont {Paredes}\ \emph {et~al.}(2005)\citenamefont {Paredes}, \citenamefont {Verstraete},\ and\ \citenamefont {Cirac}}]{PhysRevLett.95.140501}%
  \BibitemOpen
  \bibfield  {author} {\bibinfo {author} {\bibfnamefont {B.}~\bibnamefont {Paredes}}, \bibinfo {author} {\bibfnamefont {F.}~\bibnamefont {Verstraete}},\ and\ \bibinfo {author} {\bibfnamefont {J.~I.}\ \bibnamefont {Cirac}},\ }\bibfield  {title} {\bibinfo {title} {Exploiting quantum parallelism to simulate quantum random many-body systems},\ }\href {https://doi.org/10.1103/PhysRevLett.95.140501} {\bibfield  {journal} {\bibinfo  {journal} {Phys. Rev. Lett.}\ }\textbf {\bibinfo {volume} {95}},\ \bibinfo {pages} {140501} (\bibinfo {year} {2005})}\BibitemShut {NoStop}%
\bibitem [{\citenamefont {Hubig}\ and\ \citenamefont {Cirac}(2019)}]{SciPostPhys.6.3.031}%
  \BibitemOpen
  \bibfield  {author} {\bibinfo {author} {\bibfnamefont {C.}~\bibnamefont {Hubig}}\ and\ \bibinfo {author} {\bibfnamefont {J.~I.}\ \bibnamefont {Cirac}},\ }\bibfield  {title} {\bibinfo {title} {Time-dependent study of disordered models with infinite projected entangled pair states},\ }\href {https://doi.org/10.21468/SciPostPhys.6.3.031} {\bibfield  {journal} {\bibinfo  {journal} {SciPost Phys.}\ }\textbf {\bibinfo {volume} {6}},\ \bibinfo {pages} {031} (\bibinfo {year} {2019})}\BibitemShut {NoStop}%
\bibitem [{\citenamefont {Vervoort}\ \emph {et~al.}(2026)\citenamefont {Vervoort}, \citenamefont {Tang},\ and\ \citenamefont {Bultinck}}]{Vervoort:2025auj}%
  \BibitemOpen
  \bibfield  {author} {\bibinfo {author} {\bibfnamefont {K.}~\bibnamefont {Vervoort}}, \bibinfo {author} {\bibfnamefont {W.}~\bibnamefont {Tang}},\ and\ \bibinfo {author} {\bibfnamefont {N.}~\bibnamefont {Bultinck}},\ }\bibfield  {title} {\bibinfo {title} {{Extracting average properties of disordered spin chains with translationally invariant tensor networks}},\ }\href {https://doi.org/10.21468/SciPostPhysCore.9.2.040} {\bibfield  {journal} {\bibinfo  {journal} {SciPost Phys. Core}\ }\textbf {\bibinfo {volume} {9}},\ \bibinfo {pages} {040} (\bibinfo {year} {2026})},\ \Eprint {https://arxiv.org/abs/2504.21089} {arXiv:2504.21089 [cond-mat.dis-nn]} \BibitemShut {NoStop}%
\end{thebibliography}%

\end{document}